\documentclass[11pt, oneside]{article} 
\usepackage[margin=2cm]{geometry}                		
\usepackage[parfill]{parskip}    		
\usepackage{graphicx}			
\usepackage[utf8]{inputenc}
\usepackage[english]{babel} 
\usepackage{amsmath,amssymb,amsthm,bm} 
\usepackage[margin=2cm]{geometry}
\usepackage[color=yellow]{todonotes}
\usepackage{mathrsfs}
\usepackage{url}	
\usepackage{esint}
\usepackage{enumerate}
\usepackage{outlines}[enumerate]
\usepackage{framed,comment,enumerate}
\usepackage{upgreek}
\usepackage{pgfplots}
\usepackage{float}
\usepackage{tikz,tikz-cd}
\usepackage{rotating}
\usepackage{graphicx}
\usepackage{caption}
\usepackage{multicol}
\usepackage{cancel}
\usepackage{subcaption}
\usepackage[title]{appendix}
\usepackage{tikz-cd}
\usepackage{pgf, pgffor}
\usepackage[colorlinks]{hyperref}
\extrafloats{100}
\numberwithin{equation}{section}

\newtheorem{remark}{Remark}[section]

\theoremstyle{definition}
\newtheorem{definition}{Definition}[section]

\newcommand{\ob}[1]{\overline{#1}}
\newcommand{\wt}[1]{\widetilde{#1}}
\newcommand{\wh}[1]{\widehat{#1}}
\newcommand{\mb}[1]{\mathbf{#1}}
\newcommand{\mbb}[1]{\mathbb{#1}}

\newcommand{\mc}[1]{\mathcal{#1}}
\newcommand{\bs}[1]{\boldsymbol{#1}}

\newcommand{\scp}[2]{\left<#1\,,\,#2\right>}

\def\zh{{\bs{\widehat{z}}}}
\def\nh{{\bs{\widehat{n}}}}

\def\p{{\partial}}

\def\ep{{\epsilon}}
\def\rmd{{\color{red}{\rm d}}}
\def\ba{{\mathbf{a}}}
\def\bk{{\mathbf{k}}}
\def\bm{{\mathbf{m}}}
\def\bM{{\mathbf{M}}}
\def\bp{{\mathbf{p}}}

\def\bu{{\mathbf{u}}}
\def\bv{{\mathbf{v}}}
\def\bx{{\mathbf{x}}}
\def\bX{{\mathbf{X}}}

\def\p{\partial}

\pgfplotsset{compat=1.17}

\begin{document}

\title{\textbf{On the interactions between mean flows and inertial gravity waves
in the WKB approximation}}
\author{Darryl D. Holm, Ruiao Hu\footnote{Corresponding author, email: ruiao.hu15@imperial.ac.uk}, and Oliver D. Street\\
d.holm@imperial.ac.uk, ruiao.hu15@imperial.ac.uk, o.street18@imperial.ac.uk\\
Department of Mathematics, Imperial College London \\ SW7 2AZ, London, UK}

\maketitle

\begin{abstract}
We derive a Wentzel–Kramers–Brillouin (WKB) closure of the generalised Lagrangian mean (GLM) theory by using a phase-averaged Hamilton variational principle for the Euler--Boussinesq (EB) equations. Following Gjaja and Holm 1996, we consider 3D inertial gravity waves (IGWs) in the EB approximation. {\color{black}The GLM closure for WKB IGWs expresses EB wave mean flow interaction (WMFI) as WKB wave motion boosted into the reference frame of the EB equations for the Lagrangian mean transport velocity.} We provide both deterministic and stochastic closure models for GLM IGWs at leading order in 3D complex vector WKB wave asymptotics. This paper brings the Gjaja and Holm 1996 paper at leading order in wave amplitude asymptotics into an easily understood short form and proposes a stochastic generalisation of the WMFI equations for IGWs.  
\end{abstract}

\tableofcontents


\section{Introduction}

Inertial gravity waves (IGWs), also known as internal waves, comprise a classical form of wave disturbances in fluid motions under gravity that propagate in three-dimensional stratified, rotating, incompressible fluid and involve nonlinear dynamics among inertia, buoyancy, pressure gradients and Coriolis forces \cite{Vallis2017,SKV2021,Y2021}. 

\begin{figure}[!ht]
	\centering $\,$ 
	\includegraphics[width=0.65\textwidth]{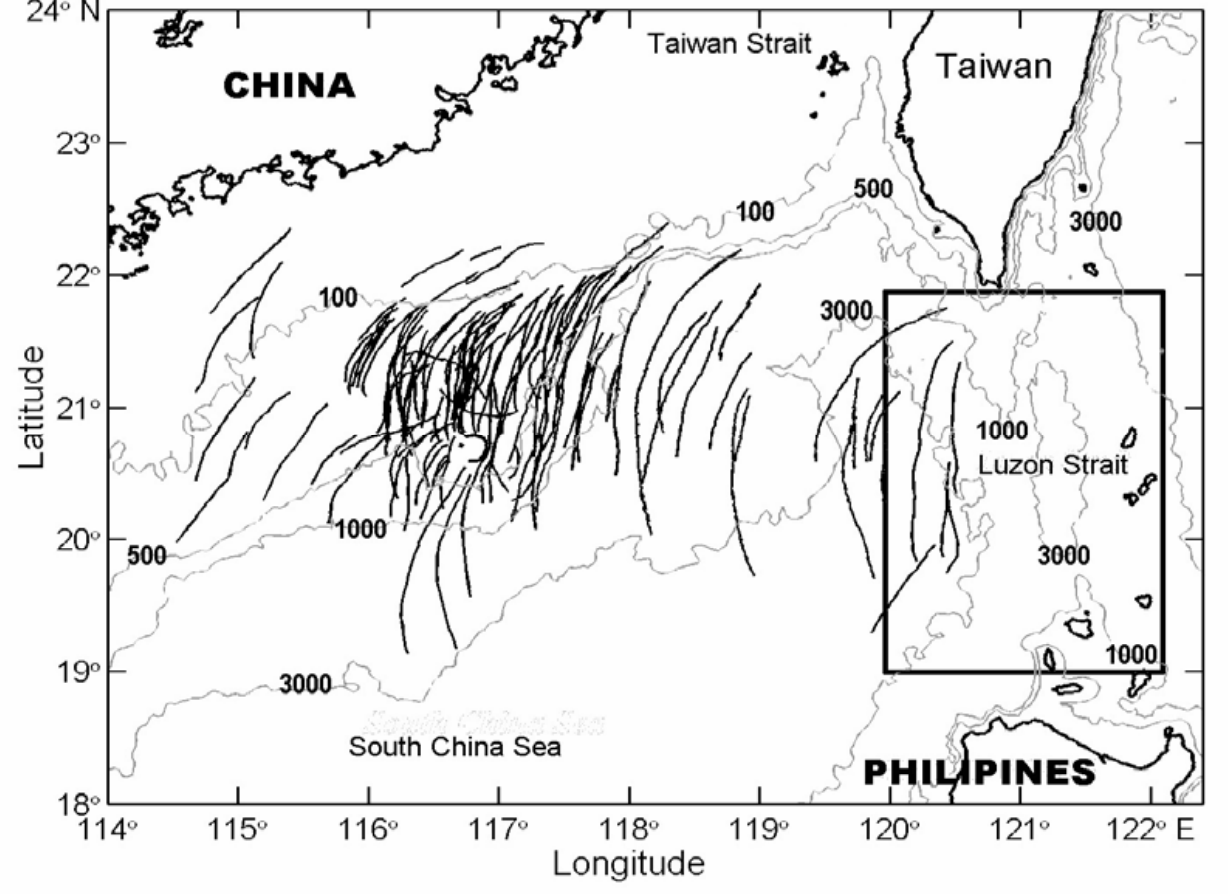}
	\caption{ The distribution of {observed IGW packets and bathymetry} in the South China Sea courtesy of \cite{ZDS2000}. Bold lines represent crest lines of leading waves in IGW packets interpreted from SAR images.  {The rectangular box on the right of this figure outlines the IGW generation source region. Looking closely near the center of this figure, one sees the crescent shape of the Dongsha atoll whose diameter is 25 {\it km}. Details of SAR images of waves near Dongsha atoll are shown in Figure \ref{fig:2}.}}
	\label{fig:1}
\end{figure}

\begin{figure}[!ht]
	\centering $\,$ 
	\includegraphics[width=0.67\textwidth]{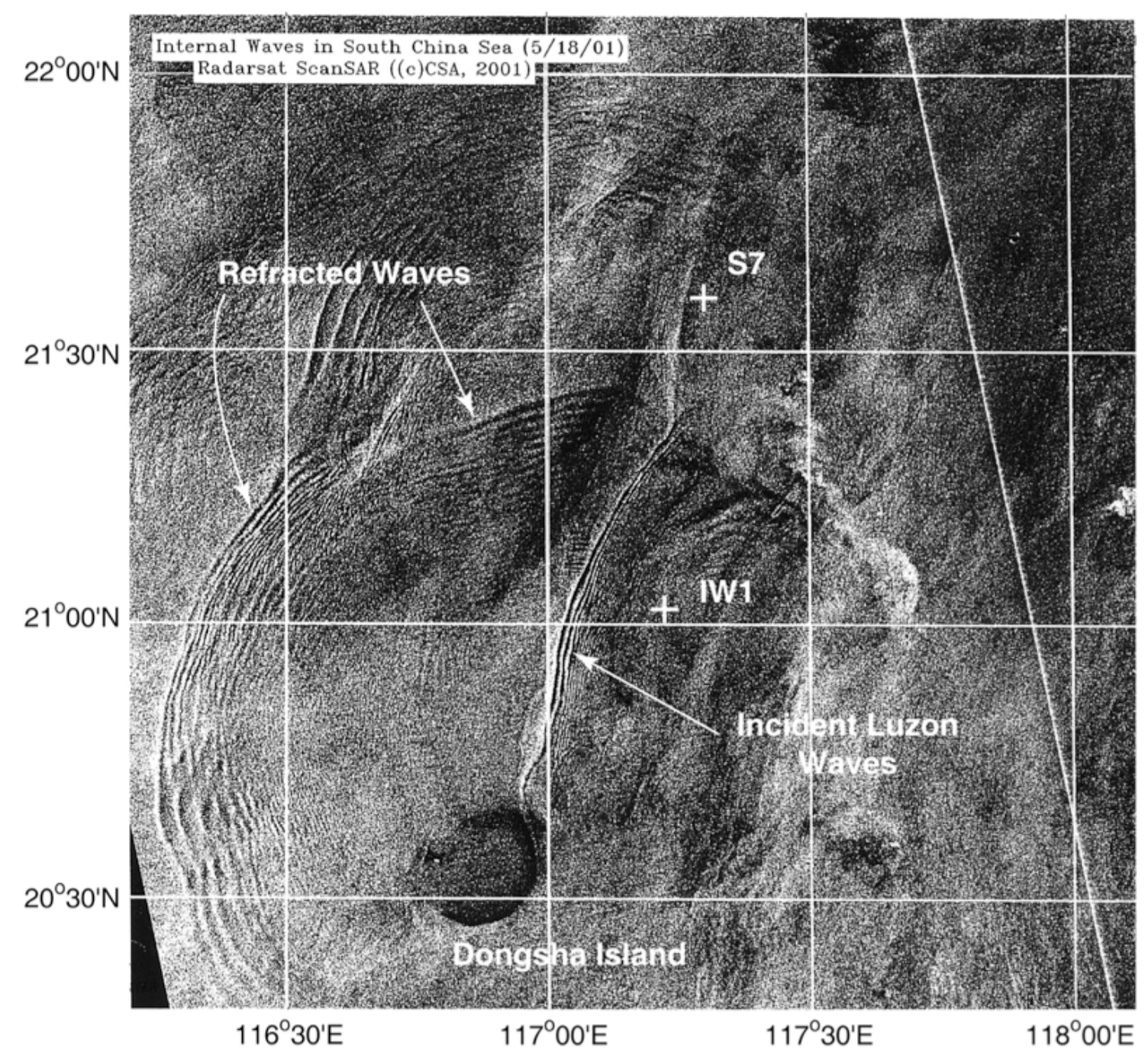}
	\caption{A satellite image showing the strong surface signatures of IGWs in the South China Sea near Dongsha atoll. Notice also the pronounced roughness of the surface due to surface gravity waves through which the IGW surface signatures propagate.  For discussion of other observations, see, e.g., \cite{HsuLiu2000}. }
	\label{fig:2}
\end{figure}

{\bf Satellite images and field data.} Satellite Synthetic Aperture Radar (SAR) is a powerful sensor for ocean remote sensing, because of its continuous capabilities and high spatial resolution. The spatial resolution of the state-of-the-art satellite SAR images reaches $20m$ –- $30 m$, and the swath width reaches $100km$ -- $450 km$. Figure \ref{fig:1} shows a typical representation of the range of SAR field data and Figure \ref{fig:2} shows a typical SAR image of IGWs on the ocean surface. 


{\bf Theoretical basis of the present work.} The paper \cite{GH1996} derived a hierarchy of approximate models of wave mean-flow interaction (WMFI) for IGWs by using asymptotic expansions and phase averages. Two different derivations of the same WMFI IGW equations were given. The first derivation was based on Fourier projections of the Euler--Boussinesq equations for a stratified rotating inviscid
incompressible fluid. The second derivation was based on Hamilton's principle for these
equations. Two small dimensionless parameters were used in the asymptotic
expansions. One small parameter was the ratio of time scales between
internal waves at most wavenumbers and the mesoscale mean flow of
the fluid. This ``adiabatic ratio'' is small and is comparable to
the corresponding ratio of space scales for the class of initial
conditions that support internal waves. The other small parameter used
in the asymptotic expansions was the ratio of the amplitude of the
internal wave to its wavelength. 
An application of Noether's theorem to the phase-averaged Hamilton's
principle showed that the resulting equations conserve the wave action,
convect a potential vorticity and can, depending on the order of
approximation, convect wave angular momentum. 
Legendre transforming from the phase-averaged
Hamilton's principle to the Hamiltonian formulation brought the WMFI theory
into the Lie-Poisson framework in which formal and nonlinear stability
analysis methods are available \cite{HMRW1985}. The Hamiltonian framework also 
revealed an analogy between the two-fluid model of
the interaction of waves and mean flow with the interaction of the superfluid
and normal fluid components of liquid $He^4$ without vortices. 
The relations to similar results for the Charney-Drazin
non-acceleration theorem, Whitham averaging, WKB stability theory, Craik-Leibovich theory of Langmuir circulations as well as the 
generalised Lagrangian-mean (GLM) fluid equations for prescribed wave displacements were also discussed in \cite{GH1996}.

{\bf Goal of the present work.} Our goal here is to derive 3D IGW equations in the class of wave mean flow interaction (WMFI) derived in  \cite{GH1996} as a mutual interaction of the mean fluid flow and the slowly varying envelope of fluctuation dynamics that is consistent with IGWs in the full 3D Euler--Boussinesq fluid flow. Physically, we take nonhydrostatic pressure effects on the wave dispersion relation into account and derive consistent nonlinear feedback effects of the internal of waves on the generation of fluid circulation based on a dynamic version of the well-known Craik-Leibovich theory of Langmuir circulation \cite{CL1976}.  Mathematically, we introduce the two WMFI degrees of freedom by factorising the full 3D Euler--Boussinesq flow map into the composition of two smooth invertible maps in Hamilton's principle for Eulerian fluid dynamics \cite{HMR1998}. 



The present work derives 3D equations for wave mean flow interaction (WMFI) as WKB wave motion boosted into the reference frame of the fluid equations for the Lagrangian mean transport velocity. The final equations derived here are consistent with traditional approaches such as Craik-Leibovih (CL) theory \cite{CL1976} except that the Stokes drift velocity in the CL formulation has its own dynamics in the present formulation. The present formulation can also be considered as a WKB closure for the GLM approach \cite{AM1978}, similar also to the oscillation centre ponderomotive closure in magnetohydrodynamics \cite{SKH1984,SKH1986}. Namely, the present formulation uses a combination of asymptotic expansion and phase resonance to close the GLM equations derived by the composition-of-maps approach and obtaining explicit formulas for the wave polarisation parameters and dispersion relation for the Doppler-shifted frequency. 

Finally, the present work also formulates stochastic equations of motion for 3D WMFI dynamics, permitting a statistical representation of the uncertainty present in observational data of geophysical flows. 




\section{Deterministic 3D Euler--Boussinesq (EB)  internal gravity waves}\label{sec:3D_EB}


\subsection{Lagrangian formulation of the WMFI equations at leading order}
{
{\bf GLM theory.}
The Generalised Lagrangian Mean (GLM) theory of wave mean flow interaction (WMFI) is derived in Andrews and McIntyre \cite{AM1978} by taking the time mean $\overline{(\,\cdot\,)}$ at a fixed position $\bx$ of the Eulerian fluid velocity, $U(\bx,t)$, shifted to a rapidly fluctuating position, $\bx^\xi:=\bx+\alpha\bs{\xi}(\mb{x},t)$ with constant scale factor $\alpha\ll1$ and zero Eulerian mean $\overline{\bs{\xi}(\mb{x},t)}=0$. The Lagrangian mean velocity $\mb{u}^L(\mb{x},t)$ at Eulerian position $\bx$ is then defined via the following calculation, 
\begin{align}
U(\bx^\xi,t):=
U(\bx+\alpha\bs{\xi}(\mb{x},t) , t)= \mb{u}^L(\mb{x},t) + \alpha\frac{d}{dt}\bs{\xi}(\mb{x},t) 
\,,\quad\hbox{where}\quad
\overline{U(\bx+\alpha\bs{\xi}(\mb{x},t) )} =: \mb{u}^L(\mb{x},t) 
\,,
\label{def: ubar}
\end{align}
with 
\begin{align}
\frac{d}{dt}\bs{\xi}(\mb{x},t)=\p_t\bs{\xi} + (\mb{u}^L\cdot\nabla) \bs{\xi} =: \bu^\ell
\,,\quad
\overline{\bu^\ell}=0
\quad\hbox{and}\quad
\overline{\mb{u}^L}=\mb{u}^L
\,.
\label{def: more}
\end{align}
Consequently, the Kelvin circulation integral for GLM in a rotating frame with constant Coriolis parameter $2\bs{\Omega}$ may be derived; see, e.g., \cite{AM1978,GH1996,H2002a,H2002b,Holm2019,H2021} and appendix \ref{Appendix:expansion} for details,
\begin{align}
\begin{split}
I_{GLM}(\mb{u}^L)  
&= \oint_{c(u^L)}  \overline{ \big( \bu(\bx^\xi,t) + \bs{\Omega}\times \bx^\xi \big)\cdot d\bx^\xi }
\\&=
\oint_{c(u^L)}   \big( \bu^L(\bx,t) + \bs{\Omega}\times \bx \big) \cdot d\mathbf{x}
+ \alpha^2\overline{\big(\bs{\Omega}\times \bs{\xi}(\bx,t) 
+ \mathbf{u}^\ell(\bx,t) \big)\cdot d\bs{\xi}(\mathbf{x},t) }
\,.
\end{split}
\label{def: GLM-Kel}
\end{align}
The Lagrangian transport velocity for GLM in \eqref{def: GLM-Kel} is indeed $\mb{u}^L$. However, the Eulerian momentum per unit mass in the integrand of the GLM circulation integral in \eqref{def: GLM-Kel} acquires an order $O(\alpha^2)$ shift due to the mean effects of the quadratic nonlinearity in the last fluctuating displacement terms in \eqref{def: GLM-Kel}. 
}

{\bf Choice of GLM closure.}
Gjaja and Holm \cite{GH1996} studied the dynamics of 3D IGWs by closing the GLM theory for the case that the fluctuation displacement $\alpha\bs{\xi}(\bx,t)$ in \eqref{def: ubar} is given by a single-frequency travelling wave $\Re(\ba(\epsilon\bx,\epsilon t)e^{i\phi(\epsilon\bx,\epsilon t)/\epsilon})$ with slowly varying complex vector amplitude $\ba(\epsilon \bx,t)$ and slowly varying, but rapid phase $\phi(\epsilon\bx,\epsilon t)/\epsilon$, with $\epsilon\ll1$; so that the time averaged Lagrangian mean of the displacement field $\alpha\bs{\xi}(\bx,t)$ would be negligible.

We choose to represent the fluctuation displacement field $\bs{\xi}(\mb{x},t)$  in the following form
\begin{align}
\bs{\xi}(\mb{x},t) = \mb{a}(\epsilon\mb{x},\epsilon t)e^{i\phi(\epsilon\mb{x},\epsilon t)/\epsilon}
+ \mb{a}^*(\epsilon\mb{x},\epsilon t)e^{-i\phi(\epsilon\mb{x},\epsilon t)/\epsilon}\,,
\label{Fluct-xi}
\end{align}
and the total pressure decomposes into 
\begin{align}
    p(\bX, t) = p_0(\bX, t) + \sum_{j\geq 1}\alpha^j\left(b_j(\epsilon\bX, \epsilon t)e^{ij\phi(\epsilon\bX,\epsilon t)/\epsilon} + b^*_j(\epsilon\bX, \epsilon t)e^{-ij\phi(\epsilon\bX,\epsilon t)/\epsilon}\right)\,.
    \label{Fluct-p}
\end{align}
Here the adiabatic parameter $\epsilon$ is defined as the ratio between space and time scales of the wave oscillations and mean flow respectively. Thus, quantities that are functions of $\bx$ and $t$, for example $\bs{\xi}(\bx, t)$, have \emph{fast} dependence on $\bx$ and $t$. Likewise, quantities which are functions of $\epsilon\bx$ and $\epsilon t$, for example $\ba(\epsilon\bx, \epsilon t)$, have \emph{slow} dependence of the space and time coordinates. Thus, in the fluctuation displacement $\bs{\xi}$ in \eqref{Fluct-xi}, the fast phase dynamics is represented by $\exp{i\phi(\epsilon\bx, \epsilon t)/\epsilon}$ which is slowly  modulated by the complex vector amplitude $\ba(\epsilon\bx, \epsilon t)$.

We will apply the GLM closure in equations \eqref{Fluct-xi} and \eqref{Fluct-p} to the 3D Euler--Boussinesq equations, which can be derived from Hamilton's principle with the following reduced Lagrangian
\begin{equation}\label{eqn:EB_action}
    0 = \delta \int_{t_0}^{t_1} \int_{\mathcal{M}} \mathscr{D}\left(\frac{1}{2}|\mb{U}|^2 + \mb{U} \cdot \bs{\Omega}\times \bX - g\varrho Z\right) + p(1- \mathscr{D})\,d^3X\,dt\,,
\end{equation}
where $\mathscr{D}d^3X=d^3x_0\in {\rm Den}(\mbb{R}^3)$ is the fluid density, $\varrho\in \mathcal{F}(\mbb{R}^3)$ is the fluid buoyancy, and $\mathcal{M}$ is the spatial domain. 
Substitution of \eqref{def: ubar}, \eqref{Fluct-xi} and \eqref{Fluct-p} into the Euler-Boussinesq Lagrangian in \eqref{eqn:EB_action} followed by asymptotic expansion in $\alpha\ll1$ and $\epsilon\ll1$ at order $O(\alpha^2)$ neglecting corrections at orders $O(\alpha^2\ep)$ and $O(\alpha^4)$ and phase averaging (i.e., keeping coefficients of resonant phase factors only) produces the following wave mean flow interaction (WMFI) closure for Hamilton's principle in Eulerian fluid variables, which splits into the sum of the average mean-flow action $\bar{L}_{MF}$ and the average wave action $\bar{L}_W$, given by \cite{GH1996} and derived in Appendix \ref{Appendix:expansion} 
as, cf. equation \eqref{eqn:action_EB_after_rearranging},

\begin{equation}\label{eqn:action_EB_approximated}
\begin{aligned}
    0 &= \delta  (S_{MF} + S_W) = \delta \int_{t_0}^{t_1} (\bar{L}_{MF} + \alpha^2 \bar{L}_W)\, dt\\
    &= \delta \int_{t_0}^{t_1}\int_{\mathcal{M}} {D}\bigg[\frac12|\bu^L|^2 
    + \bu^L\cdot\bs{\Omega}\times \bx - \rho g z  + \alpha^2\widetilde{\omega}^2|\ba|^2 
    + 2i\alpha^2\widetilde{\omega}\bs{\Omega}\cdot(\bs{a}\times\bs{a}^*) \\
    &\qquad\qquad\qquad\quad - \alpha^2i\left( b\bk\cdot\ba^* - b^*\bk\cdot\ba \right) - \alpha^2a^*_ia_j\frac{\p^2p_0}{\p x_i\p x_j} \bigg] + (1-{D})p_0 + \mathcal{O}(\alpha^2\epsilon)\, d^3x\,dt  \,.
\end{aligned}
\end{equation}




The averaged fluid quantities $\bu^L(\epsilon\bx, \epsilon t)$, $D(\epsilon\bx, \epsilon t)$ and $\rho(\epsilon\bx,\epsilon t)$ are defined to have slow dependence on $\bx$ and $t$ in the averaging procedure. 
To see the construction of slow dependence from Lagrangian labels, see section (2.1) of Gjaja and Holm \cite{GH1996}. 
In the averaged wave Lagrangian $\bar{L}_W$, the wave vector and wave frequency are defined in terms of the wave phase $\phi(\epsilon\mb{x},\epsilon t)$, as  
\begin{align}
\bk(\epsilon\mb{x},\epsilon t) := \nabla_{\epsilon \bx} \phi(\epsilon \mb{x}, \epsilon t) \quad\hbox{and}\quad \omega(\epsilon\mb{x},\epsilon t) := -\frac{\p }{\p \epsilon t}\phi(\epsilon\mb{x},\epsilon t)
\,. \label{eq:wave var def}
\end{align}
The Doppler-shifted oscillation frequency, $\wt{\omega}$, due to the coupling to the mean flow $\bu^L$ is defined through the advective time derivative $\frac{d}{d\epsilon t} := \frac{\p}{\p \epsilon t} + \mb{u}^L\cdot \nabla_{\epsilon\bx}$ and the wave phase as
\begin{align}
\wt{\omega}:= -\frac{d}{d\epsilon t} \phi = - \left(\frac{\p}{\p\epsilon t}\phi + \bu^L\cdot \nabla_{\epsilon\bx} \phi \right) = \omega - \bu^L\cdot \bk 
\,.
\label{Doppler-freq}
\end{align}
Upon introducing the Doppler-shifted oscillation $\wt{\omega}$ into $\Bar{L}_W$ in \eqref{eqn:action_EB_approximated} and pairing its definition in \eqref{Doppler-freq} with a Lagrange multiplier, $N$, one arrives at the following variational principle
\begin{equation}\label{Lag-split1}
\begin{aligned}
    0 &= \delta  (S_{MF} + S_W) = \delta \int_{t_0}^{t_1} (\bar{L}_{MF} + \alpha^2 \bar{L}_W)\, dt\\
    &= \delta \int_{t_0}^{t_1}\int_{\mathcal{M}} {D}\bigg[\frac12|\bu^L|^2 
     + \bu^L\cdot\bs{\Omega}\times \bx  - \rho g z + \alpha^2\widetilde{\omega}^2|\ba|^2 + 2i\alpha^2\widetilde{\omega}\bs{\Omega}\cdot(\ba\times\ba^*) \\
    &\qquad\qquad\qquad\quad - \alpha^2i\left( b\bk\cdot\ba^* - b^*\bk\cdot\ba \right) - \alpha^2a^*_ia_j\frac{\p^2p_0}{\p x_i\p x_j} \bigg] + (1-{D})p_0   \,d^3x\\
    &\qquad\qquad\qquad + \alpha^2 \scp{N}{-\frac{\p}{\p\epsilon t}\phi - \bu^L\cdot \nabla_{\epsilon\bx} \phi - \wt{\omega}} + \mathcal{O}(\alpha^2\ep) \, dt\,.
\end{aligned}
\end{equation}
{
Since , it may not be immediately clear how to take variations of the action \eqref{eqn:action_EB_approximated}. The inclusion of the Lagrange multiplier, $N$, imposes the relationship among the Doppler-shifted frequency $\wt{\omega}$, the Lagrangian mean velocity $\bu^L$, and the phase $\phi$, thereby facilitating
the variations. Namely, the forms of the constrained variations of the velocity field $\bu^L$ and its advected quantities, $D$ and $\rho$, are shown in \eqref{eq:EP constrain variations}. All other variables have arbitrary variations. The Euler-Poincar\'e theorem can then be applied to the variational derivatives with respect to $\bu^L$, $D$, and $\rho$, to obtain an equation for the total momentum of the system, and stationarity of the action with respect to the remaining variables implies a collection of equations for the remaining dynamics. This procedure results in a closed system of equations for both waves and mean flow, and describes their mutual interaction.
}
For Hamilton's principle of least action to apply to an asymptotically expanded action, we make use of the following definition to formalise the idea of Hamilton's principle to a given order, in the situation where the action is expanded asymptotically. 
\begin{definition}[Variational derivatives in an asymptotically expanded action.]
    When making an asymptotic expansion in Hamilton's principle, the Lagrangian in terms of any new variables, $\ell(\bu^L, {D},\rho)$ for example, becomes an infinite sum. Variational derivatives are then defined \emph{under the integral} up to some order, i.e.
    \begin{equation}
    \begin{aligned}
        0 &= \delta S = \delta\int \ell(\bu^L,{D},\rho) \,dt \\
        &=: \int \scp{\frac{\delta\ell_{\alpha^2}}{\delta \bu^L}}{\delta \bu^L} + \scp{\frac{\delta\ell_{\alpha^2}}{\delta{D}}}{\delta{D}} + \scp{\frac{\delta\ell_{\alpha^2}}{\delta \rho}}{\delta \rho} + \mathcal{O}(\alpha^2\epsilon) \,,
    \end{aligned}
    \end{equation}
    where the \emph{truncated Lagrangian}, $\ell_{\alpha^2}$, is defined as the part of the Lagrangian which corresponds to these variations
    \begin{equation*}
        \ell(\bu,{D},\rho) = \ell_{\alpha^2}(\bu, {D},\rho) + H.O.T. \,.
    \end{equation*}
    Note that we have declined to use the `big O' notation in the above equation, since $\ell_{\alpha^2}$ is defined to include all terms of order less than $\alpha^2\epsilon$ \emph{as well as} any higher order terms which produce lower order terms after integrating by parts to take variational derivatives.
\end{definition}

Hamilton's action principle \eqref{Lag-split1} yields the following variations up to order $\mathcal{O}(\alpha^2)$
\begin{align}
\begin{split}
    0 &= \delta \int^{t_1}_{t_2}\left(\Bar{L}_{MF} + \alpha^2\Bar{L}_W\right)\,dt\\
    & = \int_{t_1}^{t_2} \scp{\delta \bu^L}{{D}\bu^L + {D}\bs{\Omega}\times \bx - \alpha^2N\nabla_{\epsilon\bx}\phi} + \scp{\delta \rho}{-{D}gz} + \scp{\delta b}{-\alpha^2i\bk\cdot \ba^*} + \scp{\delta b^*}{\alpha^2i\bk\cdot \ba}  \\
    &\qquad + \scp{\delta \ba}{\alpha^2\left({D}\wt{\omega}^2\ba^* + 2i\wt{\omega}\ba^*\times \bs{\Omega} + ib^*\bk - (\ba^*\cdot\nabla)\nabla p_0\right)}\\
    &\qquad + \scp{\delta \ba^*}{\alpha^2\left({D}\wt{\omega}^2\ba + 2i\wt{\omega}\ba\times \bs{\Omega} - ib\bk - (\ba\cdot\nabla)\nabla p_0\right)}\\
    &\qquad + \scp{\delta \wt{\omega}}{2\alpha^2{D}\big(\wt{\omega}|\ba|^2 + i\bs{\Omega}\cdot(\ba\times \ba^*)\big) - \alpha^2N} + \scp{\delta {D}}{\varpi}\\
    &\qquad + \scp{\delta N}{-\frac{\p}{\p\epsilon t}\phi - \bu^L\cdot \nabla_{\epsilon\bx} \phi - \wt{\omega}} + \scp{\delta \phi}{\frac{\p}{\p\epsilon t} N + \text{div}_{\epsilon\bx}(\bu^L N) + i\text{div}_{\epsilon\bx}({D}b\ba^* -{D}b^*\ba)} \,dt\\
    &\qquad + \scp{\delta p_0}{1-{D}} + \mathcal{O}(\alpha^2\epsilon)
    \,.
    \end{split} \label{eq:Lag vars}
\end{align}
where we have 
\begin{equation}
\begin{aligned}
     \varpi :&= \delta\big(\Bar{L}_{MF} + \alpha^2\Bar{L}_W\big)/\delta D
            \\&= \frac{1}{2}|\bu^L|^2 -\rho g z + \bu^L\cdot \bs{\Omega}\times \bx - p_0 \\
     &\quad+ \alpha^2\left(\wt{\omega}^2|\ba|^2 + 2i\wt{\omega}\bs{\Omega}\cdot(\ba\times \ba^*) - i(b\bk\cdot \ba^* - b^*\bk\cdot \ba) - a^*_ia_j\frac{\p^2 p_0}{\p x_i \p x_j}\right)\,.
\end{aligned}
\label{eqn:varpi}
\end{equation}
The constrained variations in \eqref{eq:Lag vars} take the Euler-Poincar\'e form \cite{HMR1998}
\begin{align}
    \delta \bu^L &= \frac{\p}{\p\epsilon t}\bv + \bu^L\cdot\nabla_{\epsilon\bx} \bv - \bv\cdot \nabla_{\epsilon\bx}\bu^L\,, \quad \delta \rho = -\bv\cdot\nabla_{\epsilon\bx} \rho\,,\quad \delta {D} = -\text{div}_{\epsilon\bx}(\bv{D})\,,\label{eq:EP constrain variations}
\end{align}
where the $\epsilon$ appears in the derivatives of the constrained variations due to their slow dependence on space and time. Note that when isolating the arbitrary variations, $\bv$, through integration by parts, $\nabla_{\epsilon\bx}$ does not generate higher order terms when operating on $\varpi$.
From the constrained variations, one has that $\rho$ and ${D}$ are advected by the flow which then satisfies the following advection equations
\begin{align}
    \frac{\p}{\p \epsilon t} {D} + \text{div}_{\epsilon\bx}(\bu^L{D}) = 0\,,\quad \frac{\p}{\p \epsilon t} \rho + \bu^L\cdot \nabla_{\epsilon\bx} \rho = 0
    \,. \label{eq: D rho advection eq}
\end{align}
The \emph{total momentum} of the mean and fluctuating parts of the flow is defined through the variational derivative w.r.t to $\bu^L$, which is given by
\begin{align}
\mb{M} := {D}\bu^L  +{D}\bs{\Omega}\times \bx - \alpha^2N\nabla_{\epsilon\bx} \phi \,,    
\label{Def: M momentum}
\end{align}
which through the Euler-Poincar\'e theorem \cite{HMR1998}, satisifes the Euler-Poincar\'e equation \
\begin{align}
\begin{split}
    &\frac{\p}{\p\epsilon t}\left(\frac{\mb{M}}{{D}}\right) - \bu^L\times {\rm curl}_{\epsilon\bx} \left(\frac{\mb{M}}{{D}}\right) + \nabla_{\epsilon\bx}\left(\frac{1}{2}|\bu^L|^2 + p_0\right) + \frac{1}{\epsilon}g\rho \zh \\
    &\qquad \qquad + \alpha^2\nabla_{\epsilon\bx}\left( -{\omega} \frac{N}{{D}} + \wt{\omega}^2|\ba|^2 + a_ia^*_j\frac{\p^2 p_0}{\p x_i\p x_j}\right) = 0\,,
\end{split}
\label{eq:EP eq total momentum}
\end{align}
where $\zh := \nabla_{\bx} z$. Stationarity of the sum of actions $S_{MF} + \alpha^2S_W$ in \eqref{Lag-split1} under variations with respect to the fluid variables $(\bu^L,{D},\rho)$ has produced the equations for the mean flow, with order $O(\alpha^2)$ wave forcing which arises from the dependence of $\alpha^2 \bar{L}_W$ on the fluid variables ${D}$ and $\rho$. We note from the variation in $p_0$ that incompressibility of the Lagrangian mean velocity holds only within the asymptotic regime, and does not hold in an exact form. Indeed,
\begin{equation}
    {D} = 1 - \alpha^2\epsilon^2 \frac{\p^2}{\p \epsilon x_i \p \epsilon x_j} \big( {D}a^*_ia_j \big) = 1 + \mathcal{O}(\alpha^2\epsilon^2) \quad \Longrightarrow \quad  \text{div}_{\epsilon\bx}(\bu^L) = O(\alpha^2\epsilon)\,.
\label{eqn: modified D}
\end{equation}

{\bf Conservation of wave action density.}
Keeping only resonant combinations in the Lagrangian $\bar{L}_W$ in \eqref{Lag-Legendre} has eliminated any explicit dependence on the phase, $\phi$. Hence, a symmetry of the Lagrangian under constant phase shift, $\phi\to\phi+\phi_0$, has arisen. Consequently, one expects that Noether's theorem will yield a conservation law for the conjugate momentum $N$ under variations in $\phi$ of the average wave Lagrangian, $\bar{L}_W$.
The arbitrary variation $\delta \wt{\omega}$ in \eqref{eq:Lag vars} reveals the definition of $N$ as
\begin{equation}
    N := \frac{\delta \bar{L}_W}{\delta \wt{\omega}} 
    = 2{D} \big( \wt{\omega} |\ba|^2 + i \bs{\Omega}\cdot \ba\times\ba^* \big) \,,
\label{N-def}
\end{equation} 
and the arbitrary variation $\delta \phi$ in \eqref{eq:Lag vars} produces the following wave action conservation law,
\begin{equation}
\frac{\p N}{\p \epsilon t} + \text{div}_{\epsilon \bx}\Big( N\big( \bu^L  + \bv_G\big)\Big) = 0\,, 
\quad\hbox{where}\quad \bv_G:= \frac{i {D}}{ N} (\ba^* b-\ba b^*) = \frac{2 {D}}{ N}\Im(\ba b^*)
\,.\label{N-cons}
\end{equation}
Thus, the wave action $N$ is transported in an Eulerian frame by the sum of the Lagrangian mean velocity $\bu^L$ and the group velocity of the waves, $\bv_G$, defined above in \eqref{N-cons}. The evolution equation of $\phi$ in \eqref{Doppler-freq} can be written in terms of $N$ as follows
\begin{align}
    \frac{\p}{\p\epsilon t}\phi + \bu^L\cdot\nabla_{\epsilon\bx}\phi = \frac{1}{2D|\ba|^2}\left(N - 2Di\bs{\Omega}\cdot \ba\times \ba^*\right)\,, \label{eq:phi evo}
\end{align}
thus removing the explicit dependence on $\wt{\omega}$. The equations \eqref{N-cons} and \eqref{eq:phi evo} are in fact canonical Hamilton's equations boosted to the reference frame of the mean flow $\bu^L$ which is discussed in section \ref{sec:Hamiltonian structure}.

\begin{remark}[Boundary conditions for integrations by parts.]
In taking variations of wave properties, we are not considering a free upper boundary. Instead, we have set
\begin{equation}
(\nh\cdot\delta\ba^*)
\ba\cdot \frac{\p p}{\p\bx} = 0
\quad\hbox{and}\quad
\delta\phi\ \nh\cdot
N\big( \bu^L  + \bv_G\big) = 0\,,
\label{bc}
\end{equation}
on the boundary, when integrating by parts. This means that the displacement of the
wave amplitude and the flux of wave action density are both taken to be
tangential to the boundary.
\end{remark}

Combining the evolution equation of wave action density $N$ \eqref{N-cons} and wave phase $\phi$ \eqref{Doppler-freq}, one has the evolution equation of the internal wave momentum $\mb{p}/{D} := \alpha^2 N\nabla_{\epsilon\bx}\phi/{D}$.   
\begin{equation}
    \frac{\p}{\p \epsilon t} \frac{\mb{p}}{{D}} 
    - \bu^L\times \big(\nabla_{\epsilon\bx} \times \frac{\mb{p}}{{D}}\big) 
    + \nabla_{\epsilon\bx}\left(\bu^L\cdot \frac{\mb{p}}{{D}}\right) 
    = -\frac{\alpha^2}{{D}}\left( N \nabla\, \wt{\omega} 
    + \bk\, \text{div}_{\epsilon\bx}\big( N\bv_G\big) \right)
    \,. \label{eq:EP eq wave momentum}
\end{equation}
The Euler-Poincar\'e equations for the total momentum \eqref{eq:EP eq total momentum} and wave momentum \eqref{eq:EP eq wave momentum} may be assembled into the Euler-Poincar\'e equation for the mean flow momentum, $\mb{m} = {D}\bu^L + {D} \bs{\Omega}\times \bx$. 
Dividing this through by the advected mass density, ${D}$, gives the following equation for $\bu^L$
\begin{equation}
\begin{aligned}
    &\frac{\p}{\p\epsilon t}\bu^L - \bu^L\times {\rm curl}_{\epsilon\bx} \left(\bu^L 
    + \bs{\Omega}\times\bx\right) + \nabla_{\epsilon\bx}\left(\frac{1}{2}|\bu^L|^2 + p_0\right) + \frac{1}{\epsilon}g\rho \zh \\
    & \qquad =  -\alpha^2\nabla_{\epsilon\bx}\left( -\wt{\omega} \frac{N}{{D}} + \wt{\omega}^2|\ba|^2 + a_ia^*_j\frac{\p^2 p_0}{\p x_i\p x_j}\right) - \frac{\alpha^2}{{D}}\left( N \nabla_{\epsilon\bx}\, \wt{\omega} + \bk\, \text{div}_{\epsilon\bx}\big( N\bv_G\big) \right) \,. 
    \label{eq:EP eq mf momentum}
\end{aligned}
\end{equation}
{\color{black}
\begin{remark}[Hydrostatic and geostrophic balances]
As explained in section 2 of Gjaja and Holm \cite{GH1996}, at leading order $O(1/\epsilon)$ the motion equation \eqref{eq:EP eq mf momentum} establishes hydrostatic and geostrophic balances, namely
\begin{equation}
  2\bs{\Omega}\times \bu^L(\epsilon\bx,\epsilon t) + g\rho(\bx,\epsilon t) \zh + \frac{\p p_0(\bx,\epsilon t)}{\p\bx} = 0\,.
\label{eqn-balance}
\end{equation}
In order to provide the restoring force for internal waves, the advected 
relative density (or, buoyancy) $\rho\big(l^A(\bx,t)\big)$ must have one
derivative of order $O(1)$ with respect to the vertical coordinate $z$. 
In order to contribute to the wave component of the pressure gradient at order $O(\alpha^2)$ in the motion equation 
\eqref{eq:EP eq mf momentum}, the mean pressure $p_0$ must have two derivatives of order 
$O(1)$ with respect to the vertical coordinate $z$.
\end{remark}
}

\begin{remark}[Kelvin's circulation theorem for WMFI]
The two Euler-Poincar\'e equations for the total momentum $\mb{M}$ and mean flow momentum $\bm$ readily implies their respective Kelvin-circulation theorems. Namely, for the mean flow momentum $\bm$, \eqref{eq:EP eq mf momentum} implies the following 
\begin{align}
\begin{split}
    \frac{d}{d\epsilon t}\oint_{c(\bu^L)} \big(\bu^L+\bs{\Omega}\times \bx \big)\cdot d\bx
    &+ \oint_{c(\bu^L)} \frac{1}{\epsilon}\rho g\zh \cdot d\bx
\\&  =  - \,\alpha^2\oint_{c(\bu^L)} {D}^{-1}\Big(N \nabla_{\epsilon\bx} \wt{\omega}
  + \bk\, {\rm div}_{\epsilon\bx}\big( N\bv_G\big)\Big)\cdot d\bx
  \,,
\end{split}
\label{wmf-KelThm1}
\end{align}
in which one notes that the internal wave terms contribute to the creation of circulation of the mean flow at order $O(\alpha^2)$.
For the total momentum $\mb{M}$, equation \eqref{eq:EP eq total momentum} implies that
\begin{align}
    \frac{d}{d\epsilon t}\oint_{c(\bu^L)} \big(\bu^L+\bs{\Omega}\times \bx
    - \alpha^2 {D}^{-1}N\bk\big)\cdot d\bx
    + \oint_{c(\bu^L)}\frac{1}{\epsilon} \rho g\zh \cdot d\bx
    =  0  \,.
\label{wmf-KelThm2}
\end{align}
Thus, just as for the introduction of Stokes drift velocity into the integrand of Kelvin's circulation theorem in Craik-Leibovich theory \cite{CL1976}, one may regard the additional 
non-inertial force of the internal waves on the mean flow circulation as arising from a shift in the momentum per unit mass in the Kelvin circulation integrand, performed to include the internal wave degree of freedom.

\end{remark}


{\bf Legendre transforming wave Lagrangian $\bar{L}_W$ into canonical phase space variables.}
By using the definitions of $N$ and $\wt{\omega}$, one can compute the Legendre transform of $\bar{L}_W$ to obtain the following WMFI Hamiltonian $\bar{H}_W$,
\begin{align}
\label{eqn:WaveLegendreTransform}
\begin{split}
    \bar{H}_W &:= \scp{N}{\wt{\omega}} - \Bar{L}_W = \int_\mathcal{M} {D}\left(\widetilde{\omega}^2|\ba|^2 + i\left(b\bk\cdot\ba^* - b^*\bk\cdot\ba\right) + a^*_ia_j\frac{\p^2p_0}{\p x_i\p x_j}\right)\,d^3x \\
    &=\int_\mathcal{M}\frac{1}{4 {D}|\ba|^2}\left(N - 2i{D}\bs{\Omega}\cdot\ba\times \ba^*\right)^2 + i{D}\left(b\bk\cdot\ba^* - b^*\bk\cdot\ba\right) + {D} a^*_ia_j\frac{\p^2p_0}{\p x_i\p x_j}\,d^3x\,,
\end{split}
\end{align}
where we have dropped the dependence on higher order terms $O(\alpha^2\ep,\alpha^4)$ in the aymptotic expansion. Inserting the expression \eqref{eqn:WaveLegendreTransform} for $\bar{H}_W$ into \eqref{Lag-split1} yields the phase space expression of $\bar{L}_W$ as 
\begin{align}
\label{Lag-Legendre}
\begin{split}
    \Bar{L}_W = \int_\mathcal{M} -N\left(\frac{\p \phi}{\p \epsilon t}  + \bu^L \cdot \nabla_{\epsilon\bx} \phi\right) & - \frac{1}{4 {D}|\ba|^2}\left(N - 2i{D}\bs{\Omega}\cdot\ba\times \ba^*\right)^2 \\
    &- i{D}\left(b\ba^* - b^*\ba\right)\cdot\nabla\phi - {D} a^*_ia_j\frac{\p^2p_0}{\p x_i\p x_j} + \mathcal{O}(\alpha^2\epsilon) \,d^3x\,.
\end{split}
\end{align}
{
\begin{remark}[Physical interpretation of GLM WMFI]
The variations of the WKB mean wave Lagrangian $\bar{L}_W$ in the variables $N$ and $\phi$ recover canonical Hamiltonian WKB wave equations \eqref{N-cons} and \eqref{eq:phi evo} for $N$ and $\phi$. These canonical equations have been boosted into the reference frame of the Lagrangian mean transport velocity $\bu^L$. Moreover, the Lagrangian mean transport velocity $\bu^L$ satisfies the Euler-Boussinesq equations on the left-hand side of equation \eqref{wmf-KelThm1}. Thus, the phase space expression of the wave Lagrangian $\bar{L}_W$ provides the physical interpretation of the WKB mean wave motion in GLM. Namely, GLM expresses WMFI as WKB wave motion boosted into the reference frame of the Euler-Boussinesq equations satisfied by the Lagrangian mean transport velocity, $\bu^L$, and its corresponding pressure, $p_0$, and density, $\rho$. The dependence of the wave Lagrangian $\alpha^2 \bar{L}_W$ on the fluid variables ${D}$ and $\rho$ implies from its variation in $p_0$ that incompressibility of the Lagrangian mean transport velocity, $\bu^L$, no longer holds exactly. Indeed, equation \eqref{eqn: modified D} shows that the divergence of $\bu^L$ is of order $O(\alpha^2\epsilon)$, which would need to be considered when going beyond the order of asymptotics $O(\alpha^2)$ considered here. 
\end{remark}}

\begin{remark}[Potential vorticity (PV) advection theorem for WMFI]\label{PVremark}
Rewriting the indicated operations in the Kelvin circulation theorem for WMFI 
after applying the Stokes thereom gives us
\begin{align}
(\p_t + \mc{L}^\epsilon_{u_L})d\big(D^{-1}\mb{M}\cdot d\bx\big) + \frac{1}{\epsilon} g d\rho\wedge dz = 0
\,,
\label{wmf-PV}
\end{align}
where $\mc{L}^\epsilon$ denotes the Lie-derivative taken w.r.t to the rescaled basis $\epsilon\bx$. 
Since $D$ and $\rho$ are advected, i.e. they satisfies the advection equations \eqref{eq: D rho advection eq}, one finds
\begin{align}
\left(\p_t + \mc{L}^\epsilon_{u_L}\right)\Big(d\big(D^{-1}\mb{M}\cdot d\bx\big)\wedge d\rho\Big)
=
(\p_t + \mc{L}^\epsilon_{u_L})\Big(D^{-1}\nabla_{\epsilon\bx} \rho \cdot {\rm curl}_{\epsilon\bx} \big(D^{-1}\mb{M})\,D\,d^3x\big)
=0
\,.
\label{wmf-PVCONS}
\end{align}
Consequently, one finds the following total advective conservation law for WMFI {potential vorticity {PV}}, 
\begin{align}
\frac{\p}{\p\epsilon t}q + \bu^L\cdot \nabla_{\epsilon\bx} q = 0
\,,\quad\hbox{where}\quad
q:= {D}^{-1}\nabla_{\epsilon\bx} \rho \cdot {\rm curl}_{\epsilon\bx}
\big(\bu^L+\bs{\Omega}\times \bx + \alpha^2 {D}^{-1}N\bk\big) 
\,.\label{wmf-PVcons}
\end{align}
\end{remark}



{\bf Solving for wave polarisation parameters / Lagrange multipliers $p$, $b$, $b^*$, $\ba$ and $\ba^*$. }\\
The quantities $p$ and $b$ in \eqref{Lag-split1} are Lagrange multipliers which impose the incompressibility constraints for volume preservation $D=1$ and transversality of the wave vectors $\bk\cdot\ba=0$, respectively. The complex vector wave amplitudes $\ba$ and $\ba^*$ are also Lagrange multipliers whose variations impose a linear relationship among most of the wave variables. In particular, stationarity of wave action $S_W$ under variations of wave polarisation parameters $b$ and $\ba^*$ gives, respectively, 
\begin{align}
\bk\cdot\ba=0
\quad\hbox{and}\quad
\wt{\omega}^2\ba- 2i \wt{\omega} \bs{\Omega} \times \ba - (\ba\cdot \nabla) \frac{\p p_0}{\p \bx}  = ib \bk
\,,
\label{Polarisation-relations}
\end{align}
from which $b$ follows easily from the first constraint, upon taking the dot product of $\bk$ with the second constraint,
\begin{align}
    |\bk|^2 ib = - 2i\wh{\omega}(\bs{\Omega}\times \ba)\cdot \bk -\bk \cdot (\ba\cdot\nabla)\nabla p_0
    = - k^l\Big(  2i\wt{\omega}\wh{\Omega}_{lj} 
    +  (p_0)_{lj}\Big)a^j
    \,,
\label{b-eqn}
\end{align}
where $\wh{\Omega}_{ij}= - \epsilon_{ijk}\Omega^k$ and the complex vector amplitude $\ba$ is found from the $3\times3$ Hermitian matrix inversion,
\begin{align}
\begin{bmatrix}
\wt{\omega}^2 -  (p_0)_{11}  & i \wt{\omega}2\wh{\Omega}_{12} -  (p_0)_{12}  & i \wt{\omega}2\wh{\Omega}_{13} -  (p_0)_{13} \\
i\wt{\omega} 2\wh{\Omega}_{12} -  (p_0)_{12}  &  \wt{\omega}^2 -  (p_0)_{22}   & i \wt{\omega}2\wh{\Omega}_{23} -  (p_0)_{23} \\
i\wt{\omega} 2\wh{\Omega}_{13} -  (p_0)_{13}  & i \wt{\omega}2\wh{\Omega}_{23} -  (p_0)_{23}  &  \wt{\omega}^2 -  (p_0)_{33}   
\end{bmatrix}
\mathbb{P}_\perp
\begin{bmatrix}
a_1 \\ a_2 \\ a_3
\end{bmatrix}
=
ib
\begin{bmatrix}
k_1 \\ k_2 \\ k_3
\end{bmatrix},
\label{matrix-form}
\end{align}
in which the $3\times3$ matrix $\mathbb{P}_\perp$ given by 
\[
{\mathbb{P}_\perp}_{ij} := \Big(\delta_{ij}- \frac{k_ik _j}{|\bk|^2}\Big)
\]
projects out the component along $\bk$ of the complex vector amplitude $\ba\in\mbb{C}^3$. 

{\bf An index operator form of the polarisation constraints.}
The wave polarisation constraints in \eqref{Polarisation-relations} and \eqref{matrix-form} may be rewritten in index form as 
\begin{align}
a^ik_i = 0 
\quad\hbox{and}\quad
D_{ij} a^j = ibk_i 
\quad\hbox{with}\quad
D_{ij} = \wt{\omega}^2 \delta_{ij} + i\wt{\omega}2\widehat{\Omega}_{ij} 
- \frac{\p^2 p_0}{\p x^j \p x^i}
\,,\quad\hbox{so}\quad
a^{*i}D_{ij} a^j = 0
\,.
\label{Polarisation-matrix}
\end{align}
The index operator form in \eqref{Polarisation-matrix} of the polarisation relations for $(\ba,b)$ in  \eqref{Polarisation-relations} suggests a more compact representation of the wave Lagrangian, $\bar{L}_W$, as we discuss next. 

{\bf Representing the wave polarisation parameters $\ba$ and $b$ as a complex four-vector field.}
After an integration by parts using the boundary conditions in \eqref{bc}, the Eulerian action principle in \eqref{Lag-split1} may be expressed equivalently as
\begin{align}
\begin{split}
0 &= \delta (S_{MF} + \alpha^2S_W) = \delta \int_{t_0}^{t_1} (\bar{L}_{MF} + \alpha^2\bar{L}_W)\, dt
\\&:= \delta \int_{t_0}^{t_1}\int_\mathcal{M}
\bigg(\frac{D}{2}\big| \bu^L \big|^2 + D\bu^L\cdot \bs{\Omega}\times \bx - gD\rho z - p(D-1) 
\\& \hspace{3cm} 
+ \alpha^2D {F^\mu}^* D_{\mu\nu}F^\nu
+ O(\alpha^2\ep,\alpha^4)\bigg)
\,d^3x\,dt
\,,
\end{split}
\label{Lag-split2}
\end{align}

where, for notational convenience, the fields $\ba$ and $b$ comprise a complex ``four-vector field", 
\[F^\mu=(\ba,b)^T\,,\]
with $\mu  = 1, 2, 3, 4$, and the Hermitian dispersion tensor $D_{\mu\nu}=D_{\nu\mu}^*$ is given by 
\begin{align*}
D_{ij} = \wt{\omega}^2 \delta_{ij} + i\wt{\omega}2\widehat{\Omega}_{ij} 
-  \frac{\p^2 p_0}{\p x^i\p x^j}
\,,\quad 
D_{4j} = ik_j = -D_{j4}
\,,\quad 
D_{44} = 0
\,.
\end{align*}
It is clear from the decomposition of the WMFI action in \eqref{Lag-split2} that stationarity of $S_W$ with respect to variations of the fields $F=(\ba,b)^T$ yields linear relations among the wave parameters $(\ba,b)$
that recover the polarisation relations in \eqref{Polarisation-relations}
\begin{align}
D_{\mu\nu}F^\nu=0
\,.
\label{Polarisation-relations2}
\end{align}
Equation \eqref{Polarisation-relations2} recovers the linear constraints in \eqref{Polarisation-relations} on the polarization eigendirections of the field $F^\mu$ up to an overall complex constant
that can be set at the initial time.

{\bf Doppler-shifted dispersion relation.} The solvability condition $\det(D_{\mu\nu})=0$ for \eqref{Polarisation-relations2} now produces the dispersion relation for the Doppler-shifted frequency of internal gravitational waves (IGW),
\begin{equation}
\wt{\omega}^2 := (\omega - \bu^L\cdot \bk)^2 = \big(-\frac{\p}{\p \epsilon t}\phi - \bu^L\cdot\nabla_{\epsilon\bx} \phi \big)^2
= \frac{(2\bs{\Omega}\cdot\bk)^2}{|\bk|^2} + \Big(\delta_{ij}- \frac{k_ik_j}{|\bk|^2}\Big) \frac{\p^2 p_0}{\p x^i\p x^j}
\,,
\label{disp_exp}
\end{equation}
which is independent of the magnitude $|\bk|$ of the wave vector $\bk$, except for the Doppler shift due to the fluid motion. Formula \eqref{disp_exp} updates the phase $\phi$ of the wave at each time step. The complex vector amplitude $\ba$ is then found from inversion of the $3\times3$ Hermitian matrix in \eqref{matrix-form}.The remaining wave quantity $b$ is then determined from \eqref{b-eqn} at a given time step.

\begin{remark}
Under conditions of hydrostatic balance and equilibrium stratification, when $\bu^L=0$ and the pressure Hessian $p_{ij}$ has only the $p_{33}$ component,
equation \eqref{disp_exp} reduces to the well-known dispersion relation for linear internal waves
\cite{Vallis2017}. For non-equilibrium flows, though, equation \eqref{disp_exp} shows the sensitivity of the propagation of of internal waves to the pressure Hessian.
\end{remark}

\subsection{Hamiltonian structure for the WMFI equations at leading order}\label{sec:Hamiltonian structure}

Thus far, we have considered a Legendre transform within the \emph{wave} Lagrangian (see equation \eqref{eqn:WaveLegendreTransform}). It remains to perform the same calculation for the mean flow to see the full Hamiltonian structure of the model. We define the momentum of the entire flow by
\begin{equation}
    \mb{M} := {D}\bu^L + {D}\bs{\Omega}\times \bx - \alpha^2N\nabla_{\epsilon\bx}\phi =: \mb{m} - \mb{p} \,,\quad\hbox{with}\quad \mb{m} := {D}\bu^L + {D}\bs{\Omega}\times \bx\,,\quad\hbox{and}\quad \mb{p} := \alpha^2N\nabla_{\epsilon\bx}\phi \,.
\label{def: totmomentum}
\end{equation}
In the above definition, the momenta $\mb{m}$ and $\mb{p}$ are the mean and wave parts of the momentum respectively and the total momentum, $\mb{M}$, is the variational derivative of the contrained Lagrangian \eqref{Lag-split1} with respect to the Lagrangian mean velocity. We perform both the wave and mean flow Legendre transforms concurrently as
\begin{equation}
\begin{aligned}
    h  &= \scp{\mb{M}}{\bu^L} + \alpha^2\scp{N}{\omega} - \bar{L}_{MF} - \alpha^2\bar{L}_W \\
     &= \scp{D\bu^L + D\bs{\Omega}\times\bx}{\bu^L} + \alpha^2\scp{N}{\wt{\omega}}
            {- \bar{L}_{MF} - \alpha^2\bar{L}_W}\,.
\end{aligned}
\end{equation}
The resulting WMFI Hamiltonian then follows as 
\begin{align}
\begin{split}
 h{(\bM,D,\rho,\bp,N)} &= \int \Bigg\{
  \frac{1}{2{D}} \big|{\bf M} + \bp
- {D} (  \bs{\Omega}\times\bx)\big|^2
+ {D}\rho gz +
  \frac{\alpha^2 {D}}{ 4 |\ba|^2}
  \left(\frac{N}{ {D}} - 2i  \bs{\Omega}\cdot\big(\ba\times\ba^*\big)\right)^2
\\&\qquad\qquad
+ \frac{i {D}}{ N}\big(b\,\bp\cdot\ba^*-b^*\,\bp\cdot\ba\big)
  + \alpha^2{D} a^*_ia_j\frac{\p^2p_0}{\p x_i \p x_j} +
  ({D}-1)p_0 \Bigg\}
  \,d^3x \,.
  \end{split}
\label{hbar}
\end{align}
The variational derivatives of the constrained
Hamiltonian \eqref{hbar} may be determined from the
coefficients in the following expression,
\begin{align}
\begin{split}
\delta h &= \int \Bigg\{
  -\varpi \delta {D} + {D}gz\,\delta\rho
+\bu^L\cdot\delta{\bf M} -\ (1-{D})\delta p_0 + \alpha^2\Big[\wt{\omega} - \frac{i {D}}{ N}
\big(b\,\bk\cdot\ba^* - b^*\,\bk\cdot\ba\big)\Big]\,\delta N
\\&\qquad
+\ \Big[\bu^L + \bv_G\Big]\cdot\delta\bp
+\ i\alpha^2 {D} (\delta b\,\bk\cdot\ba^* - \delta b^*\,\bk\cdot\ba)
\\&\qquad
-\ \alpha^2\left[\delta\ba^*\cdot
  \left( {D}\wt{\omega}^2 \ba + 2i{D}\wt{\omega} (\bs{\Omega}\times\ba)
- i {D} b\bk - {D} \Big(\ba\cdot \frac{\p}{\p\bx}\Big)
  \frac{\p p_0}{\p\bx} \right)
+ \hbox{c.c.}\right]
\Bigg\} + \mathcal{O}(\alpha^2\epsilon) \, d^3x \,,
\end{split}
\label{vds}
\end{align}
\begin{remark}[Discussion]$\,$

\begin{itemize}    
\item
The quantity $\varpi=-\delta h/\delta D = \delta\big(\Bar{L}_{MF} + \alpha^2\Bar{L}_W\big)/\delta D$ is the Bernoulli function given by equation \eqref{eqn:varpi} now expressed as a variational partial derivative holding fixed the other variables in $h$ arising in the Legendre transform.
\item 
The Hamiltonian $h$ in \eqref{hbar} is stationary for variations in the diagnostic variables $(\ba,\ba^*,b,b^*,p_0)$. The stationary variations in these variables written in \eqref{vds} generate the constraints on the prognostic variables and the relations among the diagnostic variables. The solvability condition for these relations among the diagnostic variables determines the dispersion relation for the WKB IGWs. 
\item 
The $N$ and $\bp$ equations can combine to yield
\[
\p_{\epsilon t}\bk + \nabla_{\epsilon\bx} \omega = 0\,.
\]
{ This is the so-called `conservation of waves' equation, which imposes equality of cross derivatives of the phase function, $\phi(\epsilon \bx,\epsilon t)$.
}
\end{itemize}
\end{remark}
The above variational derivatives can be assembled into the following \emph{untangled} Lie-Poisson Hamiltonian form { which separates the dynamics of the total momentum $\bM$ in \eqref{def: totmomentum} and the advected fluid variables, $D$ and $\rho$, from the wave momentum $\bp$ and wave action density $N$},
\begin{equation}
\frac{\p}{\p \epsilon t}
\begin{bmatrix}\,
{M}_j\\ {D} \\ \rho \\ p_j \\ N
\end{bmatrix}
= - 
   \begin{bmatrix}
   {M}_k\partial_{\epsilon j} + \partial_{\epsilon k} {M}_j &{D}\partial_{\epsilon j} & - \,{\rho}_{,\epsilon j}  & 0 & 0
   \\
   \partial_{\epsilon k} {D} & 0 & 0 & 0 & 0
   \\
   {\rho}_{,\epsilon k} & 0 & 0 & 0 & 0
   \\
   0 & 0 & 0 & {p}_k\partial_{\epsilon j} + \partial_{\epsilon k} {p}_j &  N\partial_{\epsilon j} 
   \\
   0 & 0 & 0 & \partial_{\epsilon k} N & 0
   \end{bmatrix}
   \begin{bmatrix}
{\delta h/\delta {M}_k} = {u}^{L\,k}\\
{\delta h/\delta {D}} =   -\varpi  \\
{\delta h/\delta {\rho}} = {D}\,gz  \\
{\delta h/\delta {p}_k} = \big(\bu^L + \bv_G\big)^k \\
{\delta h/\delta N} =  \alpha^2\wt{\omega}
\end{bmatrix}
\,.
  \label{WMFI-LPbrkt}
\end{equation}
Here, we are using a shorthand notation for the derivatives, $\p_{\epsilon j} = \p / \p\epsilon x_j$ for example, and we have used the constraint that $\bk\cdot\bv_G = 0$ in taking the variations in $b$ and $b^*$. 

\begin{remark}\label{Remark-HamProps}
    
The \emph{untangled} Lie-Poisson Hamiltonian form in \eqref{WMFI-LPbrkt} of the ideal wave mean flow system of equations derived in the previous section from the GLM Hamilton's principle represents a constrained Lie-Poisson Hamiltonian fluid system. Its Lie-Poisson bracket is defined on the dual of the direct sum of two semidirect-product Lie algebras 
\[
\mathfrak{X}_{TOT}\circledS ({\cal F}_{MF}\oplus {\rm Den}_{MF})\oplus (\mathfrak{X}_W\circledS {\cal F}_W)\,.\]
Dual variables in $L^2(\mbb{R}^{3})$ pairing are the following, whose definitions also explain the geometric meanings of the standard calculus notation for the (MF) and (W) variables. 
\begin{itemize}
\item 
The total momentum 1-form density $\widetilde{M}=\bM\cdot d\bx\otimes d^3x\in \Lambda^1(\mbb{R}^{3})\otimes {\rm Den}(\mbb{R}^3))$ is dual to the vector fields $\mathfrak{X}_{TOT}(\mbb{R}^{3})$.
\item
The density $\widetilde{D} = Dd^3x\in {\rm Den}_{MF}(\mbb{R}^3)$ is dual to scalar functions ${\cal F}_{MF}(\mbb{R}^3)$. 
\item
The scalar function $\rho\in {\cal F}_{MF}(\mbb{R}^3)$ is dual to densities ${\rm Den}_{MF}(\mbb{R}^3)$. 
\item 
The wave momentum 1-form density $\widetilde{p}=\bp\cdot d\bx\otimes d^3x\in \Lambda^1(\mbb{R}^{3})\otimes {\rm Den}(\mbb{R}^3))$ is dual to the vector fields $\mathfrak{X}_{W}(\mbb{R}^{3})$.
\item
The wave action density $\widetilde{N} = Nd^3x$ is dual to scalar functions ${\cal F}_{W}(\mbb{R}^3)$. 
\end{itemize}

\end{remark}

\begin{remark}[Preservation of PV Casimirs]
Notice that the Casimir functions for the Hamiltonian structure of GLM WMFI in the upper left block diagonal of the Lie-Poisson operator in equation \eqref{WMFI-LPbrkt} are in the same form as for the Euler-Boussinesq fluid, except they have been modified to accommodate the wave momentum. Consequently, no Casimir functions have been gained or lost in coupling the mean flow to the fluctuations.
\end{remark}

{\bf Canonical structure of the wave dynamics.}

The wave dynamics above are written in their Lie-Poisson Hamiltonian structure. Should we return to the canonical variables, $N$ and $\phi$, then the standard canonical structure emerges. Indeed, substituting $\mb{p} = \alpha^2 N\nabla\phi$ into the Hamiltonian \eqref{hbar} and taking variations gives\footnote{The constant factor of $\alpha^2$ appearing within the canonical structure has emerged due to the choice of multiplying the constraints in Hamilton's principle by the same constant.}
\begin{align}
    \alpha^2\frac{\p \phi}{\p\epsilon t} &= -\frac{\delta h}{\delta N} = - \alpha^2\bu^L\cdot\nabla_{\epsilon\bx}\phi - \alpha^2 \widetilde{\omega}
    \label{eqn:canonical_phi}\,,\\
    \alpha^2\frac{\p N}{\p \epsilon t} &= \frac{\delta h}{\delta\phi} = -\alpha^2\,{\rm div}_{\epsilon\bx}(N\bu^L) - \alpha^2 i\,{\rm div}_{\epsilon\bx}\left( {D}(b\ba^* - b^*\ba) \right)
    \label{eqn:canonical_N}\,.
\end{align}

{\bf Tangled version of the Lie-Poisson Hamiltonian structure.}
By writing the Hamiltonian in terms of the mean flow momentum, $\mb{m}$, rather than the total momentum, $\mb{M}$, we recover the tangled version of the Lie-Poisson Hamiltonian form of the equations. Above, as in \cite{HHS2023a}, we have presented wave-current interaction in its untangled form. In a previous work \cite{HHS2023b}, the authors presented both the tangled and untangled forms, and an analogous calculation is also possible for this model of WMFI. Indeed, the Hamiltonian $h{(\bM,D,\rho,\bp,N)}$ in \eqref{hbar} becomes
\begin{align}
\begin{split}
 h'({\mb{m},D,\rho,\bp,N}) &= \int \Bigg\{ \Bigg[
  \frac{1}{2{D}} \big|\mb{m} - {D} (  \bs{\Omega}\times\bx)\big|^2
+ {D}\rho gz +
  \frac{\alpha^2 {D}}{ 4 |\ba|^2}
  \left(\frac{N}{ {D}} - 2i  \bs{\Omega}\cdot\big(\ba\times\ba^*\big)\right)^{\!2}
  \Bigg]
\\&\qquad\qquad
+ \frac{i {D}}{ N}\big(b\,\bp\cdot\ba^*-b^*\,\bp\cdot\ba\big)
  + \alpha^2{D} a^*_ia_j\frac{\p^2p_0}{\p x_i \p x_j} +
  ({D}-1)p_0 \Bigg\}
  \,d^3x 
  \,.
  \end{split}
\label{hbar_tangled}
\end{align}
The variational derivatives are largely the same, with differences only in the variation with respect to $\bp$. The {\color{black}tangled} form of the Hamiltonian equations in the Hamiltonian $h{\color{black}(\bm,D,\rho,\bp,N)}$ in \eqref{hbar_tangled} is
\begin{equation}
\frac{\p}{\p \epsilon t}
\begin{bmatrix}\,
{m}_j\\ {D} \\ \rho \\ p_j \\ N
\end{bmatrix}
= - 
   \begin{bmatrix}
   {m}_k\partial_{\epsilon j} + \partial_{\epsilon k} {m}_j & {D}\partial_{\epsilon j} & - \,{\rho}_{,\epsilon j}  & {p}_k\partial_{\epsilon j} + \partial_{\epsilon k} {p}_j & N\p_{\epsilon j}
   \\
   \partial_{\epsilon k} {D} & 0 & 0 & 0 & 0
   \\
   {\rho}_{,\epsilon k} & 0 & 0 & 0 & 0
   \\
   {p}_k\partial_{\epsilon j} + \partial_{\epsilon k} {p}_j & 0 & 0 & {p}_k\partial_{\epsilon j} + \partial_{\epsilon k} {p}_j &  N\partial_{\epsilon j} 
   \\
   \p_{\epsilon k} N & 0 & 0 & \partial_{\epsilon k} N & 0
   \end{bmatrix}
   \begin{bmatrix}
{\delta h'/\delta {m}_k} = {u}^{L\,k}\\
{\delta h'/\delta D } = -\varpi\\
{\delta h'/\delta \rho } = {D}\,gz   \\
 {\delta h'/\delta {p}_k} = \bv_G^k  \\
 {\delta h'/\delta N } = \alpha^2\wt{\omega}
\end{bmatrix} 
\,.
  \label{WMFI-LPbrkt_untangled}
\end{equation}
Instead of the direct sum in the untangled case in Remark \ref{Remark-HamProps}, this tangled Lie-Poisson bracket is defined on the dual of two nested semidirect-product Lie algebras 
\[
\big(\mathfrak{X}_{MF}\circledS ({\cal F}_{MF}\oplus {\rm Den}_{MF})\big) \,\circledS\, (\mathfrak{X}_W\circledS {\cal F}_W)\,.\] 
Corresponding dual variables in $L^2(\mbb{R}^{3})$ pairing are similar to those explained in Remark \ref{Remark-HamProps}.

\section{Stochastic WMFI}

Stochastic equations of motion may be used in fluid dynamics to model uncertainty, and such equations may be derived through Hamilton's principle \cite{H2015}. Such stochastic terms can be used to parametrise unresolved `subgridscale' dynamics absent in computational simulations, and as such are particularly relevant to geophysical applications.

Motivated by the fact that, due to computational limitations, the mean flow may only be solved for on a coarse grid when considering large scale geophysical flows, we apply the method of stochastic advection by Lie transport \cite{H2015} to the mean flow map, $\bar{g}_t$. This may be done as
\begin{equation}
    \rmd \bar{g}_tx_0 = (\bu^L\circ\bar{g}_t)x_0 \,dt + \sum_i (\bs{\zeta}_i\circ \bar{g}_t)x_0 \circ dW_t^i \,,
\end{equation}
where $W_t^i$ are independent and identically distributed Brownian motions and $\circ dW_t^i$ denotes Stratonovich integration\footnote{The notation $\circ$ may be used to denote both composition and Stratonovich integration.}. This is equivalent to
\begin{equation*}
    \rmd \bar{g}_t\bar{g}^{-1}(\bx_t) = \bu^L(\bx_t) \,dt + \sum_i \bs{\zeta}_i(\bx_t)\circ dW_t^i =: \rmd 
    \bx_t \,,
\end{equation*}
and we see that the Lagrangian mean velocity, $\bu^L$, has been stocastically perturbed. By an application of the Kunita-It\^o-Wentzell formula \cite{dLHLT2020}, we see that
\begin{equation}
    \rmd \mb{X}_t = \rmd g_tg_t^{-1}\mb{X}_t = \rmd\bx_t + \alpha^2 \big( \rmd \bs{\xi}_t(\bx_t) + \rmd\bx_t \cdot \nabla \bs{\xi}_t(\bx_t) \big) \,.
\end{equation}
Should we assume that the entire motion, corresponding to $\mb{U}_t = \dot g_t g^{-1}$, also has a stochastic part, corresponding to $\bs{\zeta}^{\xi}_i$, then we have
\begin{equation}
    \mb{U}_t\,dt + \sum_i \bs{\zeta}^{\xi}_i\circ dW_t^i = \rmd\bx_t + \alpha^2 \big( \rmd \bs{\xi}_t(\bx_t) + \rmd\bx_t \cdot \nabla \bs{\xi}_t(\bx_t) \big) \,.
\end{equation}
The uniqueness of the Doob-Meyer decomposition then indicates that each $\bs{\zeta}^{\xi}_i$ decomposes into a part corresponding to the mean flow, $\bs{\zeta}_i$, and a part corresponding to the wave motion, which we call $\bs{\sigma}_i$. 
. \cite{SC2021}.
\begin{remark}
    Following Street and Crisan, \cite{SC2021}, by the compatibility of $\bs{\xi}_t$ with the driving semimartingale, we have a representation $\rmd\bs{\xi} = \bs{A}_0\,dt + \sum_i \bs{A}_i\circ dW_t^i$. The uniqueness of the Doob-Meyer decomposition then gives  $\mb{U}_t = \bu^L + \alpha^2 \big(\bs{A}_0 + \bs{u}^L\cdot\nabla_{\epsilon\bx}\bs{\xi}_t\big)$ and $\bs{\sigma}_i = \bs{\zeta}_i + \alpha^2\big(\bs{A}_i + \bs{\zeta}_i\cdot\nabla_{\epsilon\bx}\bs{\xi}_i\big)$.
\end{remark}

WMFI is not limited to temporally averaged terms. The variability of WMFI must also be considered. This consideration results inevitably in differential equations for the slow components of the climate system, which include stochastic transport and forcing terms.
There are many ways of introducing stochasticity into the WMFI system. Some guidance in this matter can be found, e.g., in \cite{H2021}. 

In this section, we will consider two distinct framework of introducing stochasticity into Hamiltonian fluid systems. 
The first option laid out here in this section enables wave and fluid dynamics to possess different stochastically fluctuating components in their \emph{transport and phase velocities}, as follows, in which variations of the deterministic Hamiltonian below are the as those in equation \eqref{vds}. The introduction of the stochastic vector fields to the WMF evolution equations can be accomplished by making the deterministic Hamiltonian to the WMFI a semimartingale in each degree of freedom. The chosen augmentation of the Hamiltonian is based on coupling noise by $L^2$ pairings of spatially varying noise `modes' with the  momentum maps dual to the respective velocities for each degree of freedom, $\mb{m}$, $\bp$, and $N$. The coupling is done such that the variational derivatives with respect to the momentum variables will add stochastic transport terms to each of the corresponding dual velocities, as follows,
\begin{align}
\begin{split}
 \rmd h &= \int \Bigg\{ \Bigg[
  \frac{1}{2{D}} \big|\mb{m}
- {D} (  \bs{\Omega}\times\bx)\big|^2
+ {D}\rho gz +
  \frac{\alpha^2 {D}}{ 4 |\ba|^2}
  \left(\frac{N}{ {D}} - 2i  \bs{\Omega}\cdot\big(\ba\times\ba^*\big)\right)^2
  \Bigg]
\\&\qquad\qquad
+ \frac{i {D}}{ N}\big(b\,\bp\cdot\ba^*-b^*\,\bp\cdot\ba\big)
  + \alpha^2{D} a^*_ia_j\frac{\p^2p_0}{\p x_i \p x_j} \Bigg\}\,d^3x\,d\epsilon t + \int ({D}-1)\rmd p_0 \, d^3x
  \\&\qquad\qquad
  + \sum_i\int \mb{m}\cdot \bs{\zeta}_i(\bx)\circ dW_{\epsilon t}^i\,d^3x 
  + \sum_i\int \mb{p}\cdot \bs{\sigma}_i(\bx)\circ dB_{\epsilon t}^i\,d^3x
   \,.
  \end{split}
\label{hbar_stochastic}
\end{align}
where $dW_t^i$ and $dB_t^i$ are chosen to be distinct Brownian motions and $\bs{\zeta}_i(\bx)$ and $\bs{\sigma}_i(\bx)$ in principle need to be determined from calibration of transport data of each type, and leading eventually to uncertainty quantification. We have introduced a stochastic component of the pressure, thus introducing the notation $\rmd p_0$, following the framework of semimartingale driven variational principles \cite{SC2021}. The influence of the stochastic terms on the Lie-Poisson Hamiltonian dynamics can then be easily revealed, as
\begin{equation}
\rmd
\begin{bmatrix}\,
{m}_j\\ {D} \\ \rho \\ p_j \\ N
\end{bmatrix}
= - 
   \begin{bmatrix}
   {m}_k\partial_{\epsilon j} + \partial_{\epsilon k} {m}_j & {D}\partial_{\epsilon j} & - \,{\rho}_{,\epsilon j}  & {p}_k\partial_{\epsilon j} + \partial_{\epsilon k} {p}_j & N\p_{\epsilon j}
   \\
   \partial_{\epsilon k} {D} & 0 & 0 & 0 & 0
   \\
   {\rho}_{,\epsilon k} & 0 & 0 & 0 & 0
   \\
   {p}_k\partial_{\epsilon j} + \partial_{\epsilon k} {p}_j & 0 & 0 & {p}_k\partial_{\epsilon j} + \partial_{\epsilon k} {p}_j &  N\partial_{\epsilon j} 
   \\
   \p_k N & 0 & 0 & \partial_{\epsilon k} N & 0
   \end{bmatrix}
   \begin{bmatrix}
{\delta\rmd h}/{\delta m_k} = {u}^{L\,k}\,d\epsilon t + \zeta^k_i(\bx)\circ dW_{\epsilon t}^i \\
{\delta\rmd h}/{\delta D} = \ob{\pi}\,d\epsilon t + \rmd p_0\\
{\delta\rmd h}/{\delta \rho} = {D}gz \,d\epsilon t  \\
{\delta\rmd h}/{\delta p_k} =\bv_G^k \,d\epsilon t + \sigma^k_i(\bx)\circ dB_{\epsilon t}^i \\
{\delta\rmd h}/{\delta N} =\alpha^2\wt{\omega} \,d\epsilon t
\end{bmatrix} 
\,.
  \label{SALT-WMFI-LPbrkt_untangled}
\end{equation}

where $\ob{\pi}$ is given by
\[
    \ob{\pi} = -\varpi - p_0 \,,
\]
for $\varpi$ as defined in equation \eqref{eqn:varpi}. The Hamiltonian variables are as defined in the deterministic case,
\[
{\mathbf{m}}   :=  {D}({\mathbf{u}}^L + \bs{\Omega}\times \bx)
\,,\quad 
{\mathbf{p}}:= \alpha^2N\bk
\,,\quad
\bv_G:= \frac{i {D}}{ N} (\ba^* b-\ba b^*) = \frac{2 {D}}{ N}\Im(\ba b^*)
\,.\]
These variables have already appeared in the integrand of Kelvin's circulation theorem in \eqref{wmf-KelThm2}.
The stochastic version of the GLM Kelvin circulation theorem for Euler--Boussinesq incompressible flow in equation \eqref{wmf-KelThm2} thus becomes
\begin{equation}\label{GLM-comp-Kel-thm-stoch}
{\rmd}\oint_{c({\rmd}\bx_t)}
{D}^{-1} \,\mb{M} \cdot d\mathbf{x}
=
{\rmd}\oint_{c({\rmd}\bx_t)}
\Big( {\mathbf{u}}^L + \bs{\Omega}\times \bx - {D}^{-1}{\mathbf{p}}
\Big)
\cdot
d\mathbf{x}
= -\frac{1}{\epsilon}g\,
\oint_{c({\rmd}\bx_t) }
\rho \,dz\,d\epsilon t
\,,
\end{equation}
in which the material loop $c({\rmd}\bx_t)$ moves along stochastic Lagrangian trajectories given by 
the characteristics of the following stochastic vector field 
\begin{equation}\label{GLM-comp-Kel-thm-stochpath}
{\rmd}\bx_t = {\bu}^L(\bx_t,t)dt  + \sum_{a=1}^N \,\bs{\zeta}_a(\bx_t)\circ dW^a_t 
\,.
\end{equation}

{\bf A stochastic canonical structure in the wave dynamics.}

The canonical structure between the wave variables $N$ and $\phi$, noted in equations \eqref{eqn:canonical_phi} and \eqref{eqn:canonical_N}, now becomes stochastic. Indeed, substituting $\mb{M}$ and $\bp = \alpha^2N\nabla_{\epsilon\bx}\phi$ into the action and taking variations gives
\begin{align}
    \alpha^2\rmd \phi &= -\frac{\delta \rmd h}{\delta N} = - \alpha^2\bu^L\cdot\nabla_{\epsilon\bx}\phi\,d\epsilon t - \alpha^2\sum_i\bs{\zeta}_i\cdot\nabla_{\epsilon\bx}\phi \circ dW_{\epsilon t}^i - \alpha^2 \widetilde{\omega}\,d\epsilon t - \alpha^2\sum_i 
\nabla_{\epsilon\bx}\phi \cdot \bs{\sigma}_i\circ dB_{\epsilon t}^i
    \label{eqn:canonical_phi_stoch}\,,\\
    \begin{split}
    \alpha^2\rmd N &= \frac{\delta \rmd h}{\delta\phi} = -\alpha^2\,{\rm div}_{\epsilon\bx}(N\bu^L)\,d\epsilon t - \alpha^2\sum_i{\rm div}_{\epsilon\bx}(N\bs{\zeta}_i)\circ dW_{\epsilon t}^i \\
    &\qquad\qquad\qquad - \alpha^2 i\,{\rm div}_{\epsilon\bx}\left( {D}(b\ba^* - b^*\ba) \right)\,d\epsilon t - \alpha^2\sum_i\,{\rm div}_{\epsilon\bx}\left(N\bs{\sigma}_i \right)\circ dB_{\epsilon t}^i \,.
    \end{split}
    \label{eqn:canonical_N_stoch}
\end{align}
Such a stochastic generalisation of Hamilton's canonical equations has been noted and discussed for wave hydrodynamics previously \cite{S2022} for the classical water wave system.

{\bf An alternative, energy-conserving approach to the incorporation of stochastic noise.}

The second option of introducing stochasticity into the WMFI system is through the modification of mean flow and wave momentum to include different stochastically fluctuating components.
The introduction of the stochastic momentum can be accomplished by making the deterministic Lie-Poisson bracket of the WMFI system to include stochastic components.
Following \cite{HH2021}, the chosen modification is the addition of ``frozen'' Lie-Poisson bracket multiplying semi-martingales.
The fixed (frozen) parameters in the frozen Lie-Poisson brackets are the spatially, possibly temporal varying noise ``modes'' which are transformed by the deterministic transport and phase velocities in the same way as the deterministic momentum.
Let $\bs{\lambda}_i$ and $\bs{\psi}_i$ denote the stochastic fluctuations of the mean flow and wave momentum respectively, the stochastic Lie-Poisson equation can be written as
\begin{equation}
\begin{aligned}
\rmd 
\begin{bmatrix}\,
{m}_j\\ {D} \\ \rho \\ p_j \\ N
\end{bmatrix}
&= - 
   \begin{bmatrix}
   {m}_k\partial_{\epsilon j} + \partial_{\epsilon k} {m}_j & {D}\partial_{\epsilon j} & - \,{\rho}_{,\epsilon j}  & {p}_k\partial_{\epsilon j} + \partial_{\epsilon k} {p}_j & N\p_{\epsilon j}
   \\
   \partial_{\epsilon k} {D} & 0 & 0 & 0 & 0
   \\
   {\rho}_{,\epsilon k} & 0 & 0 & 0 & 0
   \\
   {p}_k\partial_{\epsilon j} + \partial_{\epsilon k} {p}_j & 0 & 0 & {p}_k\partial_{\epsilon j} + \partial_{\epsilon k} {p}_j &  N\partial_{\epsilon j} 
   \\
   \p_k N & 0 & 0 & \partial_{\epsilon k} N & 0
   \end{bmatrix}
   \begin{bmatrix}
{u}^{L\,k}\,d\epsilon t\\
-\ob{\pi} \,d\epsilon t + \rmd p_0\\
{D}\,gz \,d\epsilon t  \\
 \bv_G^k \,d\epsilon t \\
 \alpha^2\wt{\omega}\,d\epsilon t
\end{bmatrix} \\
&\qquad - \sum_i
\begin{bmatrix}
   \left(\lambda^i_k\partial_{\epsilon j} + \partial_{\epsilon k} \lambda^i_j\right)\circ dW^i_{\epsilon t} & 0 & 0 & \left(\psi^i_k\partial_{\epsilon j} + \partial_{\epsilon k} \psi^i_j\right)\circ dB^i_{\epsilon t} & 0 \\
   0 & 0 & 0 & 0 & 0 \\
   0 & 0 & 0 & 0 & 0 \\
   \left(\psi^i_k\partial_{\epsilon j} + \partial_{\epsilon k} \psi^i_j\right)\circ dB^i_{\epsilon t} & 0 & 0 & \left(\psi^i_k\partial_{\epsilon j} + \partial_{\epsilon k} \psi^i_j\right)\circ dB^i_{\epsilon t} &  0  \\
    0 & 0 & 0 & 0 & 0
\end{bmatrix}
\begin{bmatrix}
{u}^{L\,k} \\
-\ob{\pi}\,d\epsilon t + \rmd p_0\\
{D}gz  \\
\bv_G^k \\
\alpha^2\wt{\omega}
\end{bmatrix}
\,.
\end{aligned}
  \label{SFLT-WMFI-LPbrkt}
\end{equation}
Here, the stochastic component of the pressure $\rmd p$ is added as before following the semimartingale driven variational principle \cite{SC2021}. Similarly to the stochastic vector fields $\bs{\zeta}^i$ and $\bs{\sigma}^i$, we need to determine $\bs{\lambda}^i$ and $\bs{\psi}^i$ through calibration with existing data for each type of momentum.
The influence of the stochasticicty on the circulation dynamics of the mean flow and wave momentum is clear from the following modified Kelvin-circulation theorem
\begin{align}
\begin{split}
    \rmd \oint_{c(\bu^L)} \big(\bu^L+\bs{\Omega}\times \bx \big)\cdot d\bx
    &+ \oint_{c(\bu^L)} \frac{1}{\epsilon}\rho g\zh + \alpha^2 {D}^{-1}\Big(N \nabla_{\epsilon\bx} \wt{\omega}
  + \bk\, {\rm div}_{\epsilon\bx}\big( N\bv_G\big)\Big)\cdot d\bx\,d\epsilon t\\
  & +\sum_i \oint_{c(\bu^L)} D^{-1} \left(\bu^L \times \frac{\p}{\p\epsilon\bx}\times \bs{\lambda}^i - \nabla_{\epsilon\bx}\left(\bu^L\cdot\bs{\lambda}^i\right)\right)\cdot d\bx \circ dW^i_{\epsilon t} \\
  &+ \sum_i \oint_{c(\bu^L)} D^{-1} \left(\bv_G \times \frac{\p}{\p\epsilon\bx}\times \bs{\psi}^i - \nabla_{\epsilon\bx}\left(\bv_G\cdot\bs{\psi}^i\right)\right)\cdot d\bx \circ dB^i_{\epsilon t} = 0
  \,,
\end{split}
\label{SFLT-wmf-KelThm1}
\end{align}
where the loop is moving with the \emph{deterministic} velocity field.
By construction, the equation \eqref{SFLT-WMFI-LPbrkt} preserves the deterministic energy path-wise as the Poisson structure remain anti-symmetric and the variational derivative of the Hamiltonian is unchanged. However, the modification of the Poisson structure implies that the standard EB fluid Casimirs are no longer conserved. 

\section{Conclusion}

In this paper we have derived a closed system of equations for the interaction of a GLM flow with the slowly varying envelope of a WKB field of internal gravity waves (IGW) by incorporating the two approximate descriptions into Hamilton's principle. Building on the work of Gjaja and Holm \cite{GH1996}, we have shown that this approach {\color{black}boosts the canonical equations for the WKB IGW into the reference frame of the Lagrangian mean transport velocity, $\bu^L$, satisfying the Euler-Boussinesq equations on the left-hand side of equation \eqref{wmf-KelThm1}. Thus, GLM expresses WMFI as WKB wave motion boosted into the reference frame of the Euler-Boussinesq equations satisfied by the Lagrangian mean transport velocity, $\bu^L$, and its corresponding pressure, $p_0$, and density, $\rho$. The dependence of the wave Lagrangian $\alpha^2 \bar{L}_W$ on the fluid variables ${D}$ and $\rho$ implies from its variation in $p_0$ that incompressibility of the Lagrangian mean transport velocity $\bu^L$ does continue to hold for the order $O(\alpha^2)$ asymptotic expansion treated here.
} 

We have further demonstrated {\color{black}how} stochasticity in the fluid {\color{black}can} permeate through both the wave and mean flow dynamics, and that such terms can be included through the variational structure. {\color{black}Moreover, this paper has identified the nested semidirect-product Lie-Poisson structure possessed by the Hamiltonian formulation of the GLM WMFI equations. The continued preservation of the fundamental Lie algebraic structure for the Hamiltonian formulation of the stochastic GLM WMFI system implies that its data calibration and uncertainty quantification can still be treated systematically using the stochastic advection by Lie transport (SALT) approach \cite{H2015}. Future work will focus next on deriving a 2D vertical slice model for these 3D GLM WMFI equations and developing data calibration methods for the 2D vertical slice model, in order to investigate the inclusion of mean internal gravity wave effects on the responses of the stochastic Eady problem.}


\section*{Acknowledgements}
We are grateful to our friends, colleagues and collaborators for their advice and encouragement in the matters treated in this paper. 
DH especially thanks C. Cotter, F. Gay-Balmaz, I. Gjaja, J.C. McWilliams, T. S. Ratiu and C. Tronci for many insightful discussions of corresponding results similar to the ones derived here for WMFI, and in earlier work together in deriving hybrid models of complex fluids, turbulence, plasma dynamics, vertical slice models and the quantum--classical hydrodynamic description of molecules. 
DH and OS were partially supported during the present work by European Research Council (ERC) Synergy grant STUOD -- DLV-856408. RH was partially supported during the present work by EPSRC scholarship (Grant No. EP/R513052/1).

\begin{appendices}
\section{Asymptotic expansion}\label{Appendix:expansion}

This appendix fills in details of the derivations of the approximations discussed in Section \ref{sec:3D_EB}. Namely, the displacement of a fluid element from its mean trajectory is represented by
\begin{equation}
    \bX_t = \bx_t + \alpha\bs{\xi}(\bx_t,t) \,,
\end{equation}
and the associated velocity is given by
\begin{equation}
    \mb{U}_t(\bX_t) = \bu^L(\bx_t,t) + \alpha \Big( \p_t \bs{\xi} (\bx_t,t) + \bu^L\cdot\nabla_{\bx_t} \bs{\xi}(\bx_t,t) \Big) =: \bu^L(\bx_t,t) + \alpha \frac{d}{dt} \bs{\xi}(\bx_t,t) \,.
\end{equation}
The fluctuating terms are assumed to have a WKB structure, lending the pressure an associated slow/fast decomposition
\begin{align}
    \bs{\xi}(\mb{x},t) &= \mb{a}(\epsilon\mb{x},\epsilon t)e^{i\phi(\epsilon\mb{x},\epsilon t)/\epsilon}
+ \mb{a}^*(\epsilon\mb{x},\epsilon t)e^{-i\phi(\epsilon\mb{x},\epsilon t)/\epsilon}\,,
\\
    p(\bX, t) &= p_0(\bX, t) + \sum_{j\geq 1}\alpha^j\left(b_j(\epsilon\bX, \epsilon t)e^{ij\phi(\epsilon\bX,\epsilon t)/\epsilon} + b^*_j(\epsilon\bX, \epsilon t)e^{-ij\phi(\epsilon\bX,\epsilon t)/\epsilon}\right)\,.
\end{align}
Making these approximations within a fluid governed by the Euler-Boussinesq equations may be performed by substituting them into Hamilton's principle, asymptotically expanding, and truncating to leave only the leading order terms. The relevant variational principle in this case is as follows
\begin{equation}\tag{\ref{eqn:EB_action} revisited}
    0 = \delta \int_{t_0}^{t_1} \int_{\mathcal{M}} \mathscr{D}\left(\frac{1}{2}|\mb{U}|^2 + \mb{U} \cdot \bs{\Omega}\times \bX - g\varrho Z\right) + p(1-\mathscr{D})\,d^3X\,dt\,.
\end{equation}
We first note that the volume form must be written in terms of the mean basis, as
\begin{equation}
    \mathscr{D}(\bX)d^3X = \mathscr{D}^{\xi}(\bx)d^3X = \mathscr{D}^{\xi}(\bx)\mathscr{J}d^3x =: {D} d^3x \,,
\end{equation}
where $$\mathscr{J} = {\rm det}\left( \delta_{ij} + \alpha\frac{\p\xi^i}{\p x^j}\right) \,.$$ Similarly, $\varrho$ also transforms as $$ \varrho(\mb{X}) = \varrho^{\xi}(\bx) =: \rho \,.$$

Before calculating the terms featuring $\mb{U}$, note that
\begin{equation}
\begin{aligned}
    \p_t \bs{\xi} (\bx_t,t) + \bu^L\cdot\nabla_{\bx_t} \bs{\xi}(\bx_t,t) &= \epsilon\frac{\p\bs{a}}{\p\epsilon t} e^{i\phi / \epsilon} + \bs{a}i \frac{\p\phi}{\p\epsilon t}e^{i\phi / \epsilon} + \epsilon\frac{\p\bs{a}^*}{\p\epsilon t}e^{-i\phi / \epsilon} - \bs{a}^*i \frac{\p\phi}{\p\epsilon t}e^{-i\phi / \epsilon} \\
    &\qquad + \epsilon e^{i\phi / \epsilon} \bu^L\cdot \nabla_{\epsilon\bs{x}} \bs{a} + i\bs{a} e^{i\phi / \epsilon} \bu^L\cdot \nabla_{\epsilon\bs{x}} \phi \\
    &\qquad + \epsilon e^{-i\phi / \epsilon} \bu^L\cdot \nabla_{\epsilon\bs{x}} \bs{a}^* - i\bs{a}^* e^{-i\phi / \epsilon} \bu^L\cdot \nabla_{\epsilon\bs{x}} \phi \\
    &= i\bs{a}e^{i\phi/\epsilon}\left( \frac{\p\phi}{\p\epsilon t} + \bu^L\cdot\nabla_{\epsilon\bx}\phi \right) + i\bs{a}^*e^{-i\phi/\epsilon}\left(-i\frac{\p\phi}{\p\epsilon t} - \bu^L\cdot\nabla_{\epsilon\bx}\phi \right) \\
    &\qquad + \epsilon e^{i\phi/\epsilon}\left( \frac{\p\bs{a}}{\p\epsilon t} + \bu^L\cdot \nabla_{\epsilon\bs{x}} \bs{a} \right) + \epsilon e^{-i\phi/\epsilon}\left( \frac{\p\bs{a}^*}{\p\epsilon t} + \bu^L\cdot \nabla_{\epsilon\bs{x}} \bs{a}^* \right) \\
    &= -i\widetilde{\omega}\bs{a}e^{i\phi/\epsilon} + i\widetilde{\omega}\bs{a}^*e^{-i\phi/\epsilon} + \mathcal{O}(\epsilon) \,,
\end{aligned}
\end{equation}
where we define $\wt{\omega} := -\frac{d}{d\epsilon t}\phi = -\frac{\p}{\p \epsilon t}\phi + \bu^L\cdot\nabla_{\epsilon\bx}\phi$ and $\bk := \nabla_{\epsilon\bx}\phi$ as in \eqref{Doppler-freq} and \eqref{eq:wave var def}. 
The may now calculate the energy terms, beginning with kinetic energy, making use of the above relation and taking the mean\footnote{In taking the mean within the action integral, we discard the terms multiplied by rapid oscillations $\exp(\pm i\phi / \epsilon)$ and $\exp(\pm 2i\phi / \epsilon)$. These non-resonant terms are assumed to oscillate to zero under the time integral.}. Note that the following relations are true within the Lagrangian, but are expressed here in isolation.
\begin{align*}
    \frac12|\mb{U}|^2 &= \frac12 \big|\bu^L + \alpha\big(\p_t \bs{\xi} (\bx_t,t) + \bu^L\cdot\nabla_{\bx_t} \bs{\xi}(\bx_t,t)\big) \big|^2 = \frac12|\bu^L|^2 + \alpha^2|\p_t \bs{\xi} (\bx_t,t) + \bu^L\cdot\nabla_{\bx_t} \bs{\xi}(\bx_t,t)|^2 \\
    &= \frac12|\bu^L|^2+ 2\alpha^2\widetilde{\omega}^2\bs{a}\cdot\bs{a}^* + \mathcal{O}(\alpha^2\epsilon)\,.
\end{align*}
The rotation term and potential energy are
\begin{align*}
    \mb{U}\cdot \bs{\Omega} \times \bX &= \left( \bu^L + \alpha\big(\p_t \bs{\xi} (\bx_t,t) + \bu^L\cdot\nabla_{\bx_t} \bs{\xi}(\bx_t,t)\right) \cdot \bs{\Omega} \times \left( \bx + \alpha\xi(\bx,t) \right) \\
    &= \bu^L\cdot\bs{\Omega}\times \bx + \bu^L\cdot\bs{\Omega}\times(\alpha\bs{\xi}) + \alpha(\p_t \bs{\xi} + \bu^L\cdot\nabla_{\bx_t} \bs{\xi})\cdot \bs{\Omega}\times\bx + \alpha(\p_t \bs{\xi} + \bu^L\cdot\nabla_{\bx_t} \bs{\xi})\cdot \bs{\Omega}\times(\alpha\bs{\xi}) \\
    &= \bu^L\cdot\bs{\Omega}\times \bx + \alpha^2(\p_t \bs{\xi} + \bu^L\cdot\nabla_{\bx_t} \bs{\xi})\cdot \bs{\Omega}\times \bs{\xi} \\
    &= \bu^L\cdot\bs{\Omega}\times \bx + 2i\alpha^2\widetilde{\omega}\bs{\Omega}\cdot(\bs{a}\times\bs{a}^*) + \mathcal{O}(\alpha^2\epsilon) \,,\\
    g\varrho Z &= g\rho ( z + \alpha \xi_3) = g\rho z \,.
\end{align*}
Within the pressure term, we need to take care of expansion in both $p^{\xi}$ and $\mathscr{J}$. We have
\begin{equation*}
    \overline{(1-\mathscr{D}^{\xi})p^{\xi}\,d^3X} = \overline{(\mathscr{J} - {D})p^{\xi}\,d^3x} = \left(\overline{\mathscr{J}p^{\xi}} - {D}\overline{p^\xi}\right)\,d^3x 
\end{equation*}
Dealing with the terms separately, we have the expanded expression for $p^\xi$ 
\begin{align*}
\begin{split}
    p^\xi(\bx) &= p_0(\bx,t) + \alpha\frac{\p p_0}{\p x_i}\xi_i + \frac{\alpha^2}{2}\frac{\p^2 p_0}{\p x_i \p x_j}\xi_i\xi_j + \mathcal{O}(\alpha^3) \\
    & \qquad + \sum_{j\leq 1} \alpha^j \left(b_j(\epsilon\bx, \epsilon t) + \alpha\epsilon \frac{\p b_j}{\p \epsilon x_i}\xi_i + \frac{\alpha^2\epsilon^2}{2} \frac{\p^2 b_j}{\p \epsilon x_i \p \epsilon x_k}\xi_i\xi_k + \mathcal{O}(\alpha^3)\right)\cdot \\
    & \qquad \exp{\left(\frac{ij}{\epsilon}\left(\phi(\epsilon\bx,\epsilon t) + \alpha\epsilon\frac{\p \phi}{\p \epsilon x_i}\xi_i + \frac{\alpha^2\epsilon^2}{2}\frac{\p^2 \phi}{\p \epsilon x_i \p \epsilon x_k}\xi_i\xi_k + \mathcal{O}(\alpha^3) \right)\right)} + c.c. \\
    &= p_0 + \alpha\frac{\p p_0}{\p x_i}\xi_i + \frac{\alpha^2}{2}\frac{\p^2 p_0}{\p x_i \p x_j}\xi_i\xi_j + \mathcal{O}(\alpha^3)\\
    & \qquad + \sum_{j\leq 1} \alpha^j \left(b_j + \alpha\epsilon \frac{\p b_j}{\p \epsilon x_i}\xi_i + \mathcal{O}(\alpha^2)\right)\exp{\left(\frac{ij\phi}{\epsilon}\right)}\left(1 + ij\alpha\frac{\p \phi}{\p \epsilon x_i}\xi_i + \mathcal{O}(\alpha^2) \right) + c.c\\
    & = p_0 + \alpha\frac{\p p_0}{\p x_i}\xi_i + \frac{\alpha^2}{2}\frac{\p^2 p_0}{\p x_i \p x_j}\xi_i\xi_j \\
    &\qquad + \exp{\left(\frac{i\phi}{\epsilon}\right)}\left(\alpha b_1 + \alpha^2\epsilon\frac{\p b_1}{\p \epsilon x_i}\xi_i + b_1i\alpha^2\frac{\p\phi}{\p\epsilon x_i}\xi_i\right) + \exp{\left(\frac{2i\phi}{\epsilon}\right)}\alpha^2 b_2 +c.c + \mathcal{O}(\alpha^3)\,,
\end{split}
\end{align*}
which gives the phase averaged expression
\begin{align*}
    \overline{p^\xi} = p_0 + \frac{\alpha^2}{2}\frac{\p^2 p_0}{\p x_i\p x_j}\left(a_i a^*_j + a^*_i a_j\right) + \alpha^2\left(\epsilon a^*_i \frac{\p b_1}{\p \epsilon x_i} + ib_1a^*_i \frac{\p \phi}{\p\epsilon x_i} + \epsilon a_i \frac{\p b^*_1}{\p \epsilon x_i} - ib^*_1 a_i \frac{\p \phi}{\p\epsilon x_i}\right) + \mathcal{O}(\alpha^3)\,.
\end{align*}
Note that  
\begin{align*}
\mathscr{J} &= \det\left(\delta_{ij} + \alpha \frac{\p \xi_i}{\p x_j}\right) = 1 + \alpha\frac{\p \xi_i}{\p x_i} +\alpha^2\left(2\delta_{ij} - 1\right) \frac{\p \xi_i}{\p x_j} \frac{\p \xi_j}{\p x_i} + \mathcal{O}(\alpha^3)\\
& = 1 + \alpha\frac{\p \xi_i}{\p x_i} +\alpha^2\left(\frac{\p \xi_1}{\p x_1} \frac{\p \xi_2}{\p x_2} + \frac{\p \xi_3}{\p x_3} \frac{\p \xi_1}{\p x_1} + \frac{\p \xi_3}{\p x_3} \frac{\p \xi_2}{\p x_2} - \frac{\p \xi_1}{\p x_2} \frac{\p \xi_1}{\p x_2} - \frac{\p \xi_1}{\p x_3} \frac{\p \xi_3}{\p x_1} - \frac{\p \xi_2}{\p x_3} \frac{\p \xi_3}{\p x_2}\right) + \mathcal{O}(\alpha^3)\\
&= 1 + \alpha\left(\left(\frac{\p a_i}{\p x_i} + \frac{i}{\epsilon}a_i\frac{\p \phi}{\p x_i}\right)\exp{(i\phi/\epsilon)} + \left(\frac{\p a^*_i}{\p x_i} - \frac{i}{\epsilon}a^*_i\frac{\p \phi}{\p x_i}\right)\exp{(-i\phi/\epsilon)}\right) \\
& \qquad + \alpha^2\left(2\delta_{ij} - 1\right)\left(\left(\frac{\p a_i}{\p x_j} + \frac{i}{\epsilon}a_i \frac{\p \phi}{x_j}\right)\exp{(i\phi/\epsilon) + c.c}\right)\left(\left(\frac{\p a_j}{\p x_i} + \frac{i}{\epsilon}a_j \frac{\p \phi}{x_i}\right)\exp{(i\phi/\epsilon) + c.c}\right) + \mathcal{O}(\alpha^3)\,.
\end{align*}
Taking the phase average gives
\begin{align*}
\overline{\mathscr{J}} &= 1 + \alpha^2(2\delta_{ij}-1)\left(\left(\frac{\p a_i}{\p x_j} + \frac{i}{\epsilon}a_i \frac{\p \phi}{\p x_j}\right)\left(\frac{\p a^*_j}{\p x_i} - \frac{i}{\epsilon}a^*_j \frac{\p \phi}{\p x_i}\right) + c.c\right) + \mathcal{O}(\alpha^3) \\
& = 1+ i\alpha^2\frac{\p \phi}{\p \epsilon \bx}\cdot \frac{\p }{\p \bx}\times \left(\ba\times \ba^*\right) + \mathcal{O}(\alpha^3)\,, 
\end{align*}
where the last equality uses the fact that we are operating under a spatial integral and integration by parts applies. 
Then, we have
\begin{align*}
    p^\xi \mathscr{J} &=  p_0 + \alpha\frac{\p p_0}{\p x_i}\xi_i + \frac{\alpha^2}{2}\frac{\p^2 p_0}{\p x_i \p x_j}\xi_i\xi_j \\
    &\qquad + \exp{\left(\frac{i\phi}{\epsilon}\right)}\left(\alpha b_1 + \alpha^2\epsilon\frac{\p b_1}{\p \epsilon x_i}\xi_i + b_1i\alpha^2\frac{\p\phi}{\p\epsilon x_i}\xi_i\right) + \exp{\left(\frac{2i\phi}{\epsilon}\right)}\alpha^2 b_2 + c.c \\
    &\qquad + p_0 \alpha \frac{\p \xi_i}{\p x_i} + \alpha^2\frac{\p \xi_i}{\p x_i}\frac{\p p_0}{\p x_j}\xi_j + \alpha^2\exp{\left(\frac{i\phi}{\epsilon}\right)}b_1\frac{\p \xi_i}{\p x_i} + p_0\alpha^2\left(2\delta_{ij} - 1\right) \frac{\p \xi_i}{\p x_j} \frac{\p \xi_j}{\p x_i} + c.c + \mathcal{O}(\alpha^3) \,.
\end{align*}
Applying phase averaging gives
\begin{align*}
    \overline{p^\xi\mathscr{J}} &=  p_0 + \frac{\alpha^2}{2}\frac{\p^2 p_0}{\p x_i\p x_j}\left(a_i a^*_j + a^*_i a_j\right) + \alpha^2\left(\epsilon a^*_i \frac{\p b_1}{\p \epsilon x_i} + ib_1a^*_i \frac{\p \phi}{\p\epsilon x_i} + \epsilon a_i \frac{\p b^*_1}{\p \epsilon x_i} - ib^*_1 a_i \frac{\p \phi}{\p\epsilon x_i}\right)\\
    &\qquad + \alpha^2\left(\left(\frac{\p a_i}{\p x_i} + i\frac{\p \phi}{\p \epsilon x_i}a_i\right)\left(a_j^*\frac{\p p_0}{\p x_j} + b^*_1\right) + \left(\frac{\p a^*_i}{\p x_i} - i\frac{\p \phi}{\p \epsilon x_i}a^*_i\right)\left(a_j\frac{\p p_0}{\p x_j} + b_1\right)\right) \\
    &\qquad +p_0 i\alpha^2\frac{\p \phi}{\p \epsilon \bx}\cdot \frac{\p }{\p \bx}\times \left(\ba\times \ba^*\right) + \mathcal{O}(\alpha^3)\,.
\end{align*}
We may assemble these statements into the following action integral, which may be regarded as an approximation of \eqref{eqn:EB_action}.
\begin{equation}\label{eqn:action_EB_after_asymptotics}
\begin{aligned}
    S = \int_{t_0}^{t_1}\int_{\mathcal{M}} &{D}\left(\frac12|\bu^L|^2 + \alpha^2\widetilde{\omega}^2|\ba|^2 - \rho g z + \bu^L\cdot\bs{\Omega}\times \bx + 2i\alpha^2\widetilde{\omega}\bs{\Omega}\cdot(\bs{a}\times\bs{a}^*)\right) \\
    &\quad +\left( \frac{\alpha^2}{2}\frac{\p^2 p_0}{\p x_i\p x_j}\left(a_i a^*_j + a^*_i a_j\right) + \alpha^2\left(\epsilon a^*_i \frac{\p b_1}{\p \epsilon x_i} + ib_1a^*_i \frac{\p \phi}{\p\epsilon x_i} + \epsilon a_i \frac{\p b^*_1}{\p \epsilon x_i} - ib^*_1 a_i \frac{\p \phi}{\p\epsilon x_i}\right) \right)(1-{D}) \\
    &\quad+ \alpha^2\left(\left(\frac{\p a_i}{\p x_i} + i\frac{\p \phi}{\p \epsilon x_i}a_i\right)\left(a_j^*\frac{\p p_0}{\p x_j} + b^*_1\right) + \left(\frac{\p a^*_i}{\p x_i} - i\frac{\p \phi}{\p \epsilon x_i}a^*_i\right)\left(a_j\frac{\p p_0}{\p x_j} + b_1\right)\right) \\
    &\quad + p_0(1-{D}) + i\alpha^2p_0\frac{\p \phi}{\p \epsilon \bx}\cdot \frac{\p }{\p \bx}\times \left(\ba\times \ba^*\right) \, d^3x\,dt \,.
\end{aligned}
\end{equation}
We now seek to simplify this integral. Firstly, we note that the following relationships hold for the last four terms on the second row of equation \eqref{eqn:action_EB_after_asymptotics}
\begin{equation*}
    \alpha^2\left(ib_1a^*_i \frac{\p \phi}{\p\epsilon x_i} - ib^*_1 a_i \frac{\p \phi}{\p\epsilon x_i}\right) (1-{D}) + \alpha^2\left( i\frac{\p \phi}{\p \epsilon x_i}a_ib^*_1 - i\frac{\p \phi}{\p \epsilon x_i}a^*_ib_1 \right) = - \alpha^2i{D} \left( b_1\bk\cdot \ba^* - b_1^*\bk\cdot\ba \right) \,,
\end{equation*}
and
\begin{equation*}
    \alpha^2 \int_{\mathcal{M}} \left(\epsilon a^*_i \frac{\p b_1}{\p \epsilon x_i} + \epsilon a_i \frac{\p b^*_1}{\p \epsilon x_i} \right) (1-{D}) + \frac{\p a_i}{\p x_i}b_1^* + \frac{\p a^*_i}{\p x_i}b_1 \,d^3x = -\alpha^2\epsilon\int_{\mathcal{M}} {D}\left(a^*_i \frac{\p b_1}{\p \epsilon x_i} + a_i \frac{\p b^*_1}{\p \epsilon x_i} \right)\,d^3x = \mathcal{O}(\alpha^2\epsilon)\,,
\end{equation*}
after integration by parts. We have thus far involved several of the order $\alpha^2$ terms on the third line of \eqref{eqn:action_EB_after_asymptotics}. The remainder of these are handled as follows
\begin{align*}
    \alpha^2i\int_{\mathcal{M}} \frac{\p \phi}{\p \epsilon x_i}a_i a_j^*\frac{\p p_0}{\p x_j} - \frac{\p \phi}{\p \epsilon x_i}a^*_ia_j\frac{\p p_0}{\p x_j} \, d^3x &= \alpha^2i\int_{\mathcal{M}} -p_0 \frac{\p}{\p x_j}\left( \frac{\p \phi}{\p \epsilon x_i}a_i a_j^* \right) + p_0 \frac{\p}{\p x_j}\left( \frac{\p \phi}{\p \epsilon x_i}a^*_ia_j \right) \,d^3x \\
    &\hspace{-70pt}= i \int_{\mathcal{M}} -\alpha^2 p_0 \frac{\p \phi}{\p \epsilon x_i} \frac{\p}{\p x_j}\left(a_i a_j^* \right) + \alpha^2 p_0 \frac{\p \phi}{\p \epsilon x_i} \frac{\p}{\p x_j}\left( a^*_ia_j \right) \,d^3x \\
    &\hspace{-70pt}= -\int_{\mathcal{M}} i\alpha^2 p_0 \frac{\p \phi}{\p \epsilon \bx}\cdot \Big( \ba (\nabla\cdot \ba^*) - \ba^*(\nabla\cdot \ba) + (\ba^*\cdot\nabla)\ba - (\ba\cdot\nabla)\ba^* \Big)\,d^3x \\
    &\hspace{-70pt}= -\int_{\mathcal{M}} i\alpha^2p_0\frac{\p \phi}{\p \epsilon \bx}\cdot \frac{\p }{\p \bx}\times \left(\ba\times \ba^*\right) \, d^3x \,,
\end{align*}
and
\begin{align*}
    \int_{\mathcal{M}} \alpha^2\epsilon\left( \frac{\p a_i}{\p \epsilon x_i} a^*_j\frac{\p p_0}{\p x_j} + \frac{\p a^*_i}{\p \epsilon x_i} a_j\frac{\p p_0}{\p x_j} \right)\,d^3x &= \int_{\mathcal{M}} \alpha^2\epsilon\left( \frac{\p a_i}{\p \epsilon x_i} a^*_j\frac{\p p_0}{\p x_j} - a^*_i\frac{\p }{\p \epsilon x_i}\bigg( a_j\frac{\p p_0}{\p x_j}\bigg) \right)\,d^3x \\
    &= -\int_{\mathcal{M}} \alpha^2 a^*_ia_j \frac{\p^2p_0}{\p x_i\p x_j} \,d^3x = -\int_{\mathcal{M}} \frac{\alpha^2}{2}\left( a_ia^*_j + a^*_ia_j \right)\frac{\p^2p_0}{\p x_i\p x_j} \,d^3x \,.
\end{align*}
Assembling this back into the action integral \eqref{eqn:action_EB_after_asymptotics} finally yields the expression for $S$ in \eqref{eqn:action_EB_approximated},
\begin{equation}\label{eqn:action_EB_after_rearranging}
\begin{aligned}
    S = \int_{t_0}^{t_1}\int_{\mathcal{M}} &{D}\bigg[\frac12|\bu^L|^2 + \alpha^2\widetilde{\omega}^2|\ba|^2 - \rho g z + \bu^L\cdot\bs{\Omega}\times \bx + 2i\alpha^2\widetilde{\omega}\bs{\Omega}\cdot(\bs{a}\times\bs{a}^*) \\
    &\qquad - \alpha^2i\left( b\bk\cdot\ba^* - b^*\bk\cdot\ba \right) - \alpha^2a^*_ia_j\frac{\p^2p_0}{\p x_i\p x_j} \bigg] + (1-{D})p_0 + \mathcal{O}(\alpha^2\epsilon)\, d^3x\,dt  \,.
\end{aligned}
\end{equation}

\end{appendices}

\end{document}